\documentclass[lettersize,journal]{IEEEtran}
\usepackage{amsmath,amsfonts}
\usepackage[norelsize,ruled,linesnumbered,noend]{algorithm2e}
\usepackage{array}
\usepackage[caption=false,font=normalsize,labelfont=sf,textfont=sf]{subfig}
\usepackage{textcomp}
\usepackage{stfloats}
\usepackage{url}
\usepackage{verbatim}
\usepackage{graphicx}
\usepackage{cite}
\usepackage{booktabs}       
\usepackage{nicefrac}       
\usepackage{microtype}      

\usepackage{amssymb, enumerate, bm, capt-of,mathtools, accents}
\usepackage{hyperref}
\usepackage[capitalise]{cleveref}

\usepackage{relsize}
\usepackage{xcolor}
\usepackage{fp, tikz, pgfplots}
\usetikzlibrary{arrows,shapes,backgrounds,patterns,fadings,matrix,arrows,calc,
	intersections,decorations.markings,
	positioning,external,arrows.meta}
\usepgfplotslibrary{external}
\usepgfplotslibrary{fillbetween}
\usepgfplotslibrary{statistics}

\usepackage{etoolbox} 

\newtheorem{theorem}{Theorem}
\newtheorem{lemma}{Lemma}
\newtheorem{proposition}{Proposition}
\newtheorem{corollary}{Corollary}
\newtheorem{remark}{Remark}
\newtheorem{assumption}{Assumption}
\newtheorem{definition}{Definition}

\makeatletter
\newcommand{\removelatexerror}{\let\@latex@error\@gobble}
\makeatother

\makeatletter
\newcommand{\subalign}[1]{%
	\vcenter{%
		\Let@ \restore@math@cr \default@tag
		\baselineskip\fontdimen10 \scriptfont\tw@
		\advance\baselineskip\fontdimen12 \scriptfont\tw@
		\lineskip\thr@@\fontdimen8 \scriptfont\thr@@
		\lineskiplimit\lineskip
		\ialign{\hfil$\m@th\scriptstyle##$&$\m@th\scriptstyle{}##$\crcr
			#1\crcr
		}%
	}
}
\makeatother

\pgfplotsset{width=5\columnwidth /5, compat = 1.13, 
	height = 60\columnwidth /100, grid= major, 
	legend cell align = left, ticklabel style = {font=\scriptsize},
	every axis label/.append style={font=\small},
	legend style = {font={\tiny,\sffamily}},title style={yshift=-7pt, font = \small} }

\tikzset{cross/.style={cross out, draw=black, minimum size=10*(#1-\pgflinewidth), inner sep=0pt, outer sep=0pt},cross/.default={1pt}}

\newcommand{\ubar}[1]{\underaccent{\bar}{#1}}

\begin{document}

\title{Episodic Gaussian Process-Based Learning Control with Vanishing Tracking Errors}

\author{Armin Lederer,~\IEEEmembership{Graduate Student Member,~IEEE,}
Jonas Umlauft,
Sandra Hirche,~\IEEEmembership{Fellow,~IEEE,}
\thanks{Armin Lederer, Jonas Umlauft and Sandra Hirche are with the Chair of Information-oriented Control (ITR), School of Computation, Information and Technology, Technical University of Munich, 80333 Munich, Germany (email: armin.lederer, jonas.umlauft, hirche@tum.de).}
}

\markboth{Journal of \LaTeX\ Class Files,~Vol.~14, No.~8, August~2021}%
{Shell \MakeLowercase{\textit{et al.}}: A Sample Article Using IEEEtran.cls for IEEE Journals}


\maketitle

\begin{abstract}
	Due to the increasing complexity of technical systems, accurate first principle
	models can often not be obtained. Supervised machine learning can mitigate this
	issue by inferring models from measurement data. Gaussian process regression
	is particularly well suited for this purpose due to its high data-efficiency
	and its explicit uncertainty representation, which allows the derivation of 
	prediction error bounds. These error bounds have been exploited to show tracking accuracy
	guarantees for a variety of control approaches, but their direct dependency on 
	the training data is generally unclear. We address this issue by deriving a
	Bayesian prediction error bound for GP regression, which we show to 
	decay with the growth of a novel, kernel-based measure of data density. Based
	on the prediction error bound, we prove time-varying tracking accuracy guarantees
	for learned GP models used as feedback compensation of unknown nonlinearities, 
	and show to achieve vanishing tracking error with increasing 
	data density. This enables us to develop an episodic approach for learning
	Gaussian process models, such that an arbitrary tracking accuracy can be guaranteed. 
	The effectiveness of the derived theory is demonstrated in several simulations.\looseness=-1
\end{abstract}

\begin{IEEEkeywords}
	Gaussian processes, machine learning, uncertain systems, data-driven control.
\end{IEEEkeywords}

\section{Introduction}

\IEEEPARstart{F}{or} many technical systems, no or only partial first principle models are available due to their complexity or a priori unknown operating conditions. Since measurement data of such systems can typically be obtained, inferring models using supervised machine learning techniques has become increasingly popular in recent years \cite{Norgard2000}. In particular, Gaussian process (GP) regression \cite{Rasmussen2006} is a popular method since it is very data-efficient \cite{Deisenroth2015, Saveriano2017} and exhibits closed-form expressions for model updates allowing on-line learning \cite{Nguyen-Tuong2009a, Meier2016, Lederer2021b}. Moreover, GP models provide an explicit measure for prediction uncertainty, which enables the confidence-based distributed aggregation of GP models \cite{Yuan2020, Lederer2022b}, and allows to tune the behavior of control towards curiosity \cite{Buisson-Fenet2019, Capone2020c} or cautiousness \cite{Umlauft2018,Hewing2020a}.

In addition to these beneficial properties, GP regression is particularly appreciated in safety-critical control due to the existence of prediction error bounds \cite{Srinivas2012}. These bounds are typically based on the close relationship between kernel methods
and GPs \cite{Kanagawa2018}, such that the reproducing kernel Hilbert space norm induced by the GP can be used as a measure of function complexity. By combining bounds on this norm and assumptions about observation noise distributions, statistical prediction error bounds
can be derived \cite{Srinivas2012,Chowdhury2017a}. They can be efficiently computed on-line in an optimization-based fashion \cite{Scharnhorst2021}, but data-dependent closed-form expressions also exist \cite{Fiedler2021}. Moreover, they reduce to deterministic bounds when the observation noise is bounded~\cite{Maddalena2020}.

Based on the prediction error bounds for learned GP models, tracking accuracy guarantees for a large variety of control laws have been derived. This can be achieved using Lyapunov theory, e.g., for feedback linearization \cite{Umlauft2020}, computed torque control \cite{Helwa2019} and sliding mode control \cite{Lima2020}, by extending stability properties of nominal model predictive control, e.g., using continuity arguments \cite{Maiworm2021}, or robust linear control, e.g., through integral quadratic constraints \cite{Fiedler2021a}. However, these approaches suffer from the crucial drawback that accuracy guarantees are global, even though the prediction error bounds from GP models are state-dependent. Therefore accuracy guarantees can be very loose 
in cases with inhomogeneously distributed training data over the state space. In such a case, the guarantees would be dominated globally by the most conservative bound derived from the region with the fewest training data.

In general, the data dependency of such accuracy guarantees for model-based control methods has barely been analyzed in detail. While it can 
be shown for feedback linearization with event-triggered on-line learning that the tracking error vanishes 
with growing noise-free data set \cite{Umlauft2020}, similar results for noisy data do not exist. Moreover, this result is limited to feedback linearizing controllers to the best of our knowledge and does not extend to other approaches. Finally, on-line learning with GPs can be realized using suitable approximations in principle \cite{Lederer2021b}, but it remains computationally expensive, such that it is not applicable to systems with limited computational resources. The computationally less demanding approach of episodic, off-line learning has been investigated in the context of optimization-based controller tuning approaches \cite{Berkenkamp2016, Marco2017}, which can be shown to provide data-dependent performance guarantees due to the close relationship to Bayesian optimization \cite{Srinivas2012, Sui2015}. 
While these guarantees can be extended to model-based 
reinforcement learning \cite{Curi2020, Curi2021}, they strongly rely on the solved optimization problems, such that they do not generalize to a wider class of control techniques. Therefore, no guarantees and conditions for the convergence of accuracy guarantees for model-based control laws employing GP models exist to the best of our knowledge. Consequently, it is an open question how we can learn a GP model in order to ensure a desired tracking error bound with such learning-based controllers.

\subsection{Contribution and Structure}

The main contribution of this article is a novel episodic learning approach 
for GP models in order to ensure arbitrary tracking accuracy when the GP is used
to compensate unknown nonlinearities in control. Such nonlinearities can be found in a wide range of applications ranging 
from underwater vehicles, where unmodeled hydrodynamic forces due to currents can appear \cite{Fossen2011}, to physical human-robot interaction, where humans introduce generally unknown torques \cite{Yu2015}.   
For the development of this approach, we first derive an easily interpretable prediction error bound
for GPs by exploiting their Bayesian foundations.
In order to allow its straightforward computation, we 
provide probabilistic Lipschitz bounds for unknown
functions based on the GP prior. 
Based on these results, we propose a kernel-based measure to evaluate the training data
density, whose flexibility we demonstrate by exemplarily illustrating it for squared
exponential (SE), Mat\'ern class and linear kernels. Moreover, we show that prediction error 
bounds directly depend on this data density measure, which allows us to prove vanishing prediction
errors with growing data density. Based on this analysis of the GP prediction error, we derive
a novel, data density-dependent tracking error bound for 
control laws in linear systems which employ
the GP model for compensation of an unknown nonlinearity.
Finally, we extend these accuracy guarantees to establish a direct relationship with
the proposed data density measure, which allows us to develop an episodic approach
for learning a GP model ensuring a specified tracking error bound.

This article is based on our prior work \cite{Lederer2019}, which purely focuses on the derivation
of probabilistic prediction error bounds depending on the posterior variance of Gaussian processes. It
significantly extends these preliminary results by establishing a direct relationship between the training data 
density and prediction error bounds. Due to this relationship, we can bound the tracking error of linear systems with an unknown 
nonlinearity compensated by a learned model directly in terms of the data density. This allows 
us to actively generate training data for achieving arbitrary tracking accuracy in an episodic approach, while \cite{Lederer2019}
only bounds the tracking error of feedback linearizing controllers with models learned from a given data set. Therefore, we extend
the analysis framework from our prior work \cite{Lederer2019} to a design method.

The remainder of this article is structured as follows: We briefly introduce
Gaussian process regression and formalize the considered problem setting in 
\cref{sec:prob}. In \cref{sec:errorbound}, we derive a novel Bayesian 
prediction error bound for GP regression and provide methods to determine all relevant
parameters based on the prior distribution. We develop a kernel-dependent 
measure of data density and establish a straightforward relationship to the GP variance, which
allows us to investigate the asymptotic 
behavior of the error bound with increasing data set size in \cref{sec:postVar}. 
In \cref{sec:safety}, we exploit these results to derive time-varying and time-independent tracking
error guarantees, which we exploit to develop a novel episodic learning algorithm for 
ensuring arbitrary tracking accuracy. Finally, in \cref{sec:numEval}, we evaluate the developed theoretical 
framework in different simulations to demonstrate its effectiveness, before we conclude the 
paper in \cref{sec:conclusion}.

\subsection{Notation}
Vectors/matrices
are denoted by lower/upper case bold symbols, the 
$n\times n$ identity matrix by~$\bm{I}_n$, the Euclidean norm 
by~$\|\cdot\|$, and $\lambda_{\min}(\bm{A})$ and $\lambda_{\max}(\bm{A})$ the
minimum and maximum real parts of the eigenvalues of a matrix $\bm{A}$, respectively. Sets are 
denoted by upper case black board bold letters, and sets restricted to positive/non-negative 
numbers have an indexed~$+$/$+,0$, e.g.,~$\mathbb{R}_+$ for 
all positive real valued numbers. The cardinality of 
sets is denoted by~$|\cdot|$ and subsets/strict subsets are indicated by \mbox{$\subseteq/\subset$}. 
Class~$\mathcal{O}$ notation is used to provide 
asymptotic upper bounds on functions.
The ceil and floor
operator are denoted by $\lceil\cdot\rceil$ and $\lfloor\cdot\rfloor$,
respectively. The Gaussian distribution with mean $\mu\in\mathbb{R}$ and 
variance $\sigma^2\in\mathbb{R}_+$ is denoted by $\mathcal{N}(\mu,\sigma^2)$. 
A chi-squared distribution with $N$ degrees of freedom is denoted by $\chi^2_N$. 
The expectation operator $E[\cdot]$ can have an additional index to specify 
the considered random variable. Finally, a function $\alpha:\mathbb{R}_{0,+}\rightarrow\mathbb{R}_{0,+}$
is in class $\mathcal{K}_{\infty}$ if it is monotonically increasing and $\alpha(0)=0$, $\lim_{x\rightarrow\infty}\alpha(x)=\infty$.
\looseness=-1

\section{Preliminaries and Problem Setting}\label{sec:prob}

In this paper, we consider the problem of controlling linear systems perturbed by an unknown nonlinearity such that they track reference trajectories with a prescribed accuracy. In order to achieve this, we employ models learned via Gaussian process regression as compensation. Therefore, we first introduce the fundamentals of Gaussian process regression in \cref{subsec:GPR}, before we formalize the problem setting in \cref{subsec:problem}.

\subsection{Gaussian Process Regression}\label{subsec:GPR}

A Gaussian process is a stochastic process such 
that any finite number of outputs\linebreak
\mbox{$\{y_1,\ldots,y_N\} \subset\mathbb{R}$}, $N\in\mathbb{N}$,
is assigned a joint Gaussian distribution with
prior mean function~$m:\mathbb{R}^d\rightarrow \mathbb{R}$ and covariance 
defined through the kernel 
$k:\mathbb{R}^d\times\mathbb{R}^d\rightarrow\mathbb{R}$
\cite{Rasmussen2006}. Without loss of generality, we assume $m(\cdot)$ to equal $0$ in the following. 
In order to perform regression with Gaussian processes, 
they are considered as a a prior distribution. This allows to employ Bayes' theorem to 
calculate the posterior distribution given a training data set $\mathbb{D}=\{(\bm{x}^{(n)},y^{(n)} \}_{n=1}^N$ 
consisting of $N$ inputs $\bm{x}^{(n)}\in\mathbb{R}^d$ and targets $y^{(n)}\in\mathbb{R}$, 
which are Gaussian perturbed measurements of an unknown function $f:\mathbb{R}^d\rightarrow\mathbb{R}$,
i.e., $y^{(n)}=f(\bm{x}^{(n)})+\epsilon^{(n)}$, $\epsilon^{(n)}\sim\mathcal{N}(0,\sigma_{\mathrm{on}}^2)$, $\sigma_{\mathrm{on}}^2\in\mathbb{R}_+$. 
Due to the properties of Gaussian distributions, the posterior is again a Gaussian process, 
which yields the posterior mean $\mu(\cdot)$ and variance $\sigma^2(\cdot)$ functions
\begin{align}\label{eq:mean}
\mu(\bm{x})&=\bm{k}^T(\bm{x})\left( \bm{K}\!+\!\sigma_{\mathrm{on}}^2\bm{I}_{N}\right)^{-1}\bm{y},\\
\sigma^2(\bm{x})&=k(\bm{x},\bm{x})\!-\!
\bm{k}^T(\bm{x})\left(\bm{K}\!+\!\sigma_{\mathrm{on}}^2\bm{I}_{N}\right)^{-1}\bm{k}(\bm{x}),
\label{eq:var}
\end{align}
where we define the kernel matrix~$\bm{K}$ and the 
kernel vector~$\bm{k}(\bm{x})$ through~$K_{ij}=k(\bm{x}^{(i)},\bm{x}^{(j)})$ and~$k_{i}(\bm{x})=k(\bm{x},\bm{x}^{(i)})$, respectively, with~$i,j=1,\ldots,N$, and~$\bm{y} = [y^{(1)} \cdots y^{(N)}]^T$.

\subsection{Problem Formulation}\label{subsec:problem}

We consider single-input linear dynamical systems with nonlinear input perturbation of the form
\begin{align}\label{eq:sys}
\dot{\bm{x}}=\bm{A}\bm{x}+\bm{b}(u+f(\bm{x}))
\end{align}
with initial condition $\bm{x}(0)=\bm{x}_0\in \mathbb{X} \subseteq \mathbb{R}^{d}$ and scalar control input $u:\mathbb{R}_{0,+}\rightarrow\mathbb{U}\subseteq \mathbb{R}$.
The matrix $\bm{A}\in\mathbb{R}^{d\times d}$ and vector $\bm{b}\in\mathbb{R}^{d}$ are assumed to be known, while we consider $f:\mathbb{X}\rightarrow\mathbb{R}$ to be an unknown nonlinearity. This system structure covers a wide range of practical systems and can represent, e.g., systems controlled via approximate feedback linearization \cite{Umlauft2020} or backstepping controllers for certain classes of dynamics \cite{Capone2019}. 
Note that we merely consider the restriction to single-input systems for notational convenience, but our derived results can be easily generalized to multi-input dynamics.

The considered task is to track a bounded reference trajectory $\bm{x}_{\mathrm{ref}}:\mathbb{R}_{0,+}\rightarrow\mathbb{R}^{d}$ with the state $\bm{x}(t)$. 
In order to enable the accurate tracking of the reference trajectory $\bm{x}_{\mathrm{ref}}(\cdot)$, we restrict ourselves to references of the form
\begin{align}\label{eq:xref_structure}
\dot{\bm{x}}_{\mathrm{ref}}=\bm{A}\bm{x}_{\mathrm{ref}}+\bm{b}r_{\mathrm{ref}},
\end{align}
where $r_{\mathrm{ref}}:\mathbb{R}_{0,+}\rightarrow\mathbb{R}$ is a reference signal. For tracking the reference trajectory, we can employ a control law
\begin{align}\label{eq:ctrl}
u = \bm{\theta}^T(\bm{x}-\bm{x}_{\mathrm{ref}})+r_{\mathrm{ref}}-\hat{f}(\bm{x}),
\end{align}
where $\bm{\theta}\in\mathbb{R}^{d}$ is a control gain vector and $\hat{f}:\mathbb{X}\rightarrow\mathbb{R}$ is a model of the unknown nonlinear perturbation $f(\cdot)$. This control law leads to closed-loop dynamics of the tracking error $\bm{e}(t)=\bm{x}(t)-\bm{x}_{\mathrm{ref}}(t)$ given by
\begin{align}\label{eq:err_dyn}
\dot{\bm{e}}=\bm{A}_{\bm{\theta}}\bm{e} + \bm{b}(f(\bm{x})-\hat{f}(\bm{x})),
\end{align}
where $\bm{A}_{\bm{\theta}}=\bm{A}-\bm{b}\bm{\theta}^T$. In order to ensure the stability of these dynamics in the case of exact model knowledge $f(\bm{x})=\hat{f}(\bm{x})$, we employ the following assumption on $\bm{A}_{\bm{\theta}}$.
\begin{assumption}\label{as:stab}
	The matrix $\bm{A}_{\bm{\theta}}$ has distinct and non-positive eigenvalues, which decrease monotonically with the parameters $\bm{\theta}$, i.e., there exists a class $\mathcal{K}_\infty$ function $\alpha:\mathbb{R}_{0.+}\rightarrow\mathbb{R}_{0,+}$ such that $\lambda_{\max}(\bm{A}_{\bm{\theta}})\leq-\alpha(\|\bm{\theta}\|)$.
\end{assumption}
This assumption essentially requires the controllability of the pair $(\bm{A},\bm{b})$ \cite{Skogestad2005}, which allows the eigenvalues of the matrix $\bm{A}_{\bm{\theta}}$ to be considered as design parameters, e.g., using methods such as pole placement. Since controllability is a common requirement in linear systems theory, \cref{as:stab} is not restrictive. Note that the requirement of distinct eigenvalues is only required to simplify the presentation in the following sections by ensuring diagonalizability of $\bm{A}_{\bm{\theta}}$, but can be avoided by generalizing the derivations using Jordan blocks \cite{Perko2006}.

While \cref{as:stab} ensures that the error dynamics \eqref{eq:err_dyn} do not diverge, the tracking precision crucially relies on the accuracy of the model $\hat{f}(\cdot)$. Therefore,  
we assume to learn it from measurements $(\bm{x}^{(n)},y^{(n)})$ using Gaussian process regression, such
that we can use $\hat{f}(\bm{x})=\mu(\bm{x})$ in the control law \eqref{eq:ctrl}. Since this merely leads
to an approximate compensation of the nonlinearity, exact tracking cannot be ensured in general. Therefore, we 
consider the problem of learning a Gaussian process model of $f(\cdot)$,
such that the tracking error is guaranteed to be probabilistically bounded by a prescribed constant $\bar{e}\in\mathbb{R}_+$, i.e., 
\begin{align}\label{eq:track_bound_prob}
\mathbb{P}\left(\|\bm{x}(t)-\bm{x}_{\mathrm{ref}}(t)\|\leq \bar{e}, ~\forall t\geq 0\right)\geq 1-\delta
\end{align}
for $\delta\in(0,1)$. 
Due to the complexity of this problem, we decompose it into the subproblems of deriving a probabilistic error bound for Gaussian process regression, analyzing the dependency of the error bounds on the training data density, and developing an approach for generating training data with sufficiently high density, such that the prescribed tracking error bound $\bar{e}$ is satisfied. These subproblems are described in more detail in the following.

\subsubsection{Probabilistic Regression Error Bounds}\label{subsubsec:error_bound}

In order to be able to ensure any bound for the tracking error $\bm{x}-\bm{x}_{\mathrm{ref}}$, it is necessary 
to find an upper bound for the learning error $f(\bm{x}(t))-\mu(\bm{x}(t))$ along the system trajectory $\bm{x}(t)$. 
Since we do not know the exact system trajectory $\bm{x}(t)$ in advance, we consider the problem of bounding the regression error in a compact domain $\mathbb{X}\subset\mathbb{R}^d$. Since the bound must hold jointly for all states $\bm{x}$ in the domain $\mathbb{X}$, we refer to it as probabilistic uniform error bound, which is formally defined as follows.
\begin{definition}\label{def:unibound}
	Gaussian process regression exhibits a uniformly bounded prediction error on a compact set $\mathbb{X}\subset\mathbb{R}^d$ with probability $1-\delta$ if there exists a function $\eta:\mathbb{X}\rightarrow\mathbb{R}_{0,+}$ such that
	\begin{align}\label{eq:unibound}
	P\left( |f(\bm{x})-\mu(\bm{x})|\leq \eta(\bm{x}),~\forall\bm{x}\in\mathbb{X} \right)\geq 1-\delta.
	\end{align}
\end{definition}
In general, we cannot expect to guarantee a uniformly bounded regression error without any regularity assumptions about the unknown function $f(\cdot)$. Due to the Bayesian foundation of 
Gaussian processes, we employ their prior distribution for this purpose, which we formalize in the following assumption. 
\begin{assumption}
	\label{ass:samplefun}
	The unknown function~$f(\cdot)$ is a sample from the 
	Gaussian process $\mathcal{GP}(0,k(\bm{x},\bm{x}'))$.
\end{assumption}
This assumption, which has similarly been used in, e.g., \cite{Dhiman2021, Srinivas2012}, has a twofold implication. On the one hand, it specifies the admissible functions for regression via the space of sample functions, which depends on the employed kernel $k(\cdot,\cdot)$. For example, it is straightforward to see that polynomial kernels can be used to learn polynomial functions of the same degree. Moreover, it is well known that the sample space of GPs with squared exponential kernel contains all continuous functions \cite{VanderVaart2011}. Therefore, choosing a suitable kernel for ensuring that the unknown function lies in the space of sample functions is usually not a challenging problem in practice. On the other hand, \cref{ass:samplefun} induces a weighting between possible sample functions due to the Gaussian process probability density. Since we base the derivation of the uniform error bound on this weighting, an unknown function $f(\cdot)$ with  low prior probability density would lead to sets $\{f'(\cdot): |f'(\bm{x})-\mu(\bm{x})|\leq\eta(\bm{x}) \}$ with a high probability under the GP prior, even though they do not contain the unknown function $f(\cdot)$. 
Hence, the true function $f(\cdot)$ should have a high probability density under the GP prior. This can be efficiently achieved in practice using suitable kernel tuning methods, e.g., \cite{Capone2021b}, or via a re-calibration of the probability distribution after training \cite{Kuleshov2018}. Therefore, ensuring 
a suitable prior distribution is not a severe limitation, such that \cref{ass:samplefun} is not restrictive in 
practice.

\subsubsection{Dependency of Error Bounds on Data Density}\label{subsubsec:e2d}

After a probabilistic uniform error bound $\eta(\cdot)$ has been derived, we consider the problem of deriving conditions for the training data $\mathbb{D}$ which ensure that the error bound $\eta(\cdot)$ stays below a desired value $\bar{\eta}\in\mathbb{R}_+$. This requires the design of a suitable measure of data density $\rho:\mathbb{X}\rightarrow\mathbb{R}_+$, which reflects the dependency of the error bound $\eta(\cdot)$ on the data distribution. Therefore, the measure $\rho(\cdot)$ must consider the information structure of the GP induced by the employed kernel $k(\cdot,\cdot)$. 

Based on the derived density measure $\rho(\cdot)$, the problem of ensuring a learning error bound $\bar{\eta}$ reduces to showing that the existence of a lower bound $\ubar{\rho}\in\mathbb{R}_+$ for the data density $\rho(\cdot)$ leads to the implication
\begin{align}
	\rho(\bm{x})\geq \ubar{\rho}\qquad \Rightarrow\qquad \eta(\bm{x})\leq \bar{\eta}(\ubar{\rho}).
\end{align}
As we want to be able to ensure arbitrary small learning error bounds $\bar{\eta}(\ubar{\rho})$, it must additionally hold that
\begin{align}\label{eq:van_pred_error}
	\lim\limits_{\ubar{\rho}\rightarrow\infty} \bar{\eta}(\ubar{\rho})=0.
\end{align}

\subsubsection{Data Generation for Guaranteed Tracking Accuracy}\label{subsubsec:dataGen}

Finally, we consider the problem of developing an episodic approach for training data generation, which 
achieves the necessary data density $\rho(\cdot)$ to ensure the satisfaction of the tracking error bound 
\eqref{eq:track_bound_prob}. Firstly, this requires the derivation of a tracking error bound, such that for
a given learning error bound $\bar{\eta}$, we have
\begin{align}
	\!\!\eta(\bm{x}_{\mathrm{ref}}(t))\leq \bar{\eta}~ \Rightarrow~ \mathbb{P}\left(\|\bm{x}(t)\!-\!\bm{x}_{\mathrm{ref}}(t)\|\leq \bar{\upsilon}(\bar{\eta})\right)\geq 1-\delta\!\!
\end{align}
for some function $\bar{\upsilon}:\mathbb{R}_{0,+}\rightarrow\mathbb{R}_{0,+}$. 
Similarly as in \eqref{eq:van_pred_error}, this bound must also vanish asymptotically, i.e., 
\begin{align}
	\lim\limits_{\bar{\eta}\rightarrow 0}\bar{\upsilon}(\bar{\eta}) = 0,
\end{align}
in order to admit arbitrarily small tracking error guarantees. 

Using this tracking error bound and the derived dependency of the learning error bound $\eta(\cdot)$ on the data density $\rho(\cdot)$, the problem of developing a data generation approach simplifies to finding an episodic roll-out strategy satisfying
\begin{align}\label{eq:rho_episodic}
	\ubar{\rho}_{i+1}>\ubar{\rho}_i,\qquad\qquad\qquad \lim\limits_{i\rightarrow\infty}\ubar{\rho}_i = \infty,
\end{align}
where the index $i$ is used to denote the roll-out episode. This ensures that there exists a finite number of episodes $N_E\in\mathbb{N}$ such that $\bar{\upsilon}(\bar{\eta}(\ubar{\rho}_{N_E}))\leq \bar{e}$. Therefore, finding a roll-out strategy ensuring \eqref{eq:rho_episodic} solves the overall problem of learning a Gaussian process model of $f(\cdot)$ such that a prescribed error bound $\bar{e}$ is satisfied.

\section{Probabilistic Uniform Error Bound}
\label{sec:errorbound}

In this section, we derive an easily computable uniform 
error bound for Gaussian process regression based on the prior distribution addressing 
the problem described in \cref{subsubsec:error_bound}. 
We first present the uniform error bound and approaches to compute its 
parameters in \cref{subsec:regerror}. Since the bound also relies on the Lipschitz
constant of the unknown function, which is not always known a priori, we show
how a probabilistic Lipschitz constant can be derived from the prior Gaussian process
distribution in \cref{subsec:prob Lipschitz}.

\subsection{Uniform Error Bound based on Lipschitz Continuity}
\label{subsec:regerror}

Since the prior Gaussian process induces a probability distribution for each point in a compact 
set $\mathbb{X}$, we can discretize this set and exploit standard tail bounds for Gaussian 
distributions to obtain point-wise error bounds \cite{Srinivas2012}. If all involved functions are
continuous, we can straightforwardly extend these point-wise guarantees yielding the uniform
error bound presented in the following.
\begin{theorem}
	\label{th:errbound_with}
	Consider a zero mean prior Gaussian process defined
	on a compact set $\mathbb{X}$ and let $f:\mathbb{X}\rightarrow \mathbb{R}$ be a continuous unknown function with Lipschitz constant $L_f$ which satisfies \cref{ass:samplefun}. Assume the GP posterior mean  $\mu(\cdot)$ and standard deviation $\sigma(\cdot)$ 
	are continuous with Lipschitz constant $L_{\mu}$ and modulus of continuity $\omega_{\sigma}(\cdot)$.
	Moreover, pick $\delta\in (0,1)$, $\tau\in\mathbb{R}_+$ and set 
	\begin{align}
	\label{eq:beta}
	\beta_{\mathbb{X}}(\tau)&=2\log\left(\frac{M(\tau,\mathbb{X})}{\delta}\right),\\
	\gamma(\tau)&=\left( L_{\mu}+L_f\right)\tau+\sqrt{\beta_{\mathbb{X}}(\tau)}\omega_{\sigma}(\tau),
	\label{eq:gamma}
	\end{align}
	where $M(\tau,\mathbb{X})$ denotes the $\tau$-covering number of $\mathbb{X}$\footnote{The $\tau$-covering number of a set $\mathbb{X}$ is the smallest number, such there exists a set $\mathbb{X}_{\tau}$ 
	satisfying $|\mathbb{X}_{\tau}|=M(\tau,\mathbb{X})$ and $\forall\bm{x}\in\mathbb{X}$ there exists 
	$\bm{x}'\in\mathbb{X}_{\tau}$ with $\|\bm{x}-\bm{x}'\|\leq \tau$.}.
	Then, the prediction error is uniformly bounded with probability of at least $1-\delta$ on $\mathbb{X}$ 
	with bound
	\begin{align}
	\label{eq:errorbound}
	\eta(\bm{x})=\sqrt{\beta_{\mathbb{X}}(\tau)}\sigma(\bm{x})\!+\!\gamma(\tau).
	\end{align}
\end{theorem}
\begin{IEEEproof}
  We exploit the continuity properties of the posterior mean, variance and the 
   unknown function  to prove
	the probabilistic uniform error bound by exploiting the fact that for every 
	grid $\mathbb{X}_{\tau}$ with $|\mathbb{X}_{\tau}|$ grid points and 
	\begin{align}
	\max\limits_{\bm{x}\in\mathbb{X}} \min\limits_{\bm{x}'\in\mathbb{X}_{\tau}}
	\|\bm{x}-\bm{x}'\|\leq \tau
	\label{eq:gridconstant}
	\end{align}
	it holds with probability of at least
	$1-|\mathbb{X}_{\tau}|\mathrm{e}^{-\beta_{\mathbb{X}}(\tau)/2}$ that \cite{Srinivas2012}
	\begin{align}
	|f(\bm{x})-\mu(\bm{x})|\leq \sqrt{\beta_{\mathbb{X}}(\tau)}\sigma(\bm{x}) 
	\quad \forall\bm{x}\in \mathbb{X}_{\tau}.
	\end{align}
	Choose \mbox{$\beta_{\mathbb{X}}(\tau)=2\log\left(\frac{|\mathbb{X}_{\tau}|}{\delta}\right)$}, 
	then
	\begin{align}
	|f(\bm{x})-\mu(\bm{x})|\leq \sqrt{\beta_{\mathbb{X}}(\tau)}\sigma(\bm{x}) 
	\quad \forall\bm{x}\in \mathbb{X}_{\tau}
	\end{align}
	holds with probability of at least $1-\delta$. Due to continuity of 
	$f(\bm{x})$, $\mu(\bm{x})$ and $\sigma(\bm{x})$ we obtain
	\begin{align}
	\min\limits_{\bm{x}'\in\mathbb{X}_{\tau}}|f(\bm{x})-f(\bm{x}')|&\leq 
	\tau L_f\quad \forall \bm{x}\in\mathbb{X}\\
	\min\limits_{\bm{x}'\in\mathbb{X}_{\tau}}|\mu(\bm{x})-\mu(\bm{x}')|&\leq 
	\tau L_{\mu}\quad \forall \bm{x}\in\mathbb{X}\\
	\min\limits_{\bm{x}'\in\mathbb{X}_{\tau}}|\sigma(\bm{x})-\sigma(\bm{x}')|&\leq 
	\omega_{\sigma}(\tau)\quad \forall \bm{x}\in\mathbb{X}.
	\end{align}
	Moreover, the minimum number of grid points satisfying \eqref{eq:gridconstant} is 
	given by the covering number $M(\tau,\mathbb{X})$. Hence, we obtain
	\begin{align}
	P\left(\!|f(\bm{x})\!-\!\mu(\bm{x})|\!\leq\! 
	\sqrt{\beta_{\mathbb{X}}(\tau)}\sigma(\bm{x})\!+\!\gamma(\tau), 
	~\forall\bm{x}\!\in\!\mathbb{X}\!\right)\!\geq\! 1\!-\!\delta,
	\end{align}
	for $\beta_{\mathbb{X}}(\tau)$ and $\gamma(\tau)$ defined in \eqref{eq:beta} and \eqref{eq:gamma}, respectively.
\end{IEEEproof}

\begin{figure}
	\center
	\tikzsetnextfilename{cover_illustration}
	\begin{minipage}{0.78\columnwidth}
		\center
		\scalebox{0.75}{
		\begin{tikzpicture}		
		\draw[smooth cycle, tension=2, fill=black!20] plot coordinates{(0,0) (1.5,-0.98) (2.1,-0.6) (2.7,-0.7) (4.5,0.1) (3.6,0.4) (4.24,0.8) (4.8,1.4) (3.9,1.7) (3.3,1.4) (2.1,1.6)};
		
		\draw[blue, thick] (-0.1,-2.1)--(4.9,-2.1)--(4.9,2.9)--(-0.1,2.9)--(-0.1,-2.1);
		
		\fill[red] (0.73,2.35) circle (0.05cm);
		\fill[red] (2.4,2.35) circle (0.05cm);
		\fill[red] (4.07,2.35) circle (0.05cm);
		
		\fill[red] (-0.1,0.8) circle (0.05cm);
		\fill[red] (1.565,0.8) circle (0.05cm);
		\fill[red] (3.28,0.8) circle (0.05cm);
		\fill[red] (4.9,0.8) circle (0.05cm);
		
		\fill[red] (0.73,-0.75) circle (0.05cm);
		\fill[red] (2.4,-0.75) circle (0.05cm);
		\fill[red] (4.07,-0.75) circle (0.05cm);
		
		\fill[red] (-0.1,-2.1) circle (0.05cm);
		\fill[red] (1.565,-2.1) circle (0.05cm);
		\fill[red] (3.28,-2.1) circle (0.05cm);
		\fill[red] (4.9,-2.1) circle (0.05cm);
		
		\draw[red] (0.73,2.35) circle (1);
		\draw[red] (2.4,2.35) circle (1);
		\draw[red] (4.07,2.35) circle (1);
		
		\draw[red] (-0.1,0.8) circle (1);
		\draw[red] (1.565,0.8) circle (1);
		\draw[red] (3.28,0.8) circle (1);
		\draw[red] (4.9,0.8) circle (1);
		
		\draw[red] (0.73,-0.75) circle (1);
		\draw[red] (2.4,-0.75) circle (1);
		\draw[red] (4.07,-0.75) circle (1);
		
		\draw[red] (-0.1,-2.1) circle (1);
		\draw[red] (1.565,-2.1) circle (1);
		\draw[red] (3.28,-2.1) circle (1);
		\draw[red] (4.9,-2.1) circle (1);
		
		\draw[red,very thick] (4.9,-2.1)--(5.9,-2.1);
		
		\node[red] at (5.4,-2.3) {$\tau$};
		\draw [thick,decorate, decoration={brace,amplitude=17pt,mirror,raise=1pt}, yshift=0pt]
		(4.9,2.9) -- (-0.1,2.9) node [black,midway,xshift=0.0cm,yshift=0.9cm] {
			$r$};
		\end{tikzpicture}
	}
	\end{minipage}
	\begin{minipage}{0.2\columnwidth}
		\begin{tikzpicture}
			\begin{axis}[%
			hide axis,
			xmin=-5,
			xmax=5,
			ymin=0,
			ymax=1,
			height=2cm
			]
				\addlegendimage{red,only marks};
				\addlegendentry{$\mathbb{X}_{\tau}$};
				\addlegendimage{fill=black!20, area legend};
				\addlegendentry{$\mathbb{X}$};
				\addlegendimage{blue, thick};
				\addlegendentry{$\tilde{\mathbb{X}}$};
			\end{axis}
	   \end{tikzpicture}
	\end{minipage}	
	\caption{Illustration of the derivation of an upper bound for the covering number~$M(\tau,\mathbb{X})$.}
	\label{fig:cover}
\end{figure}
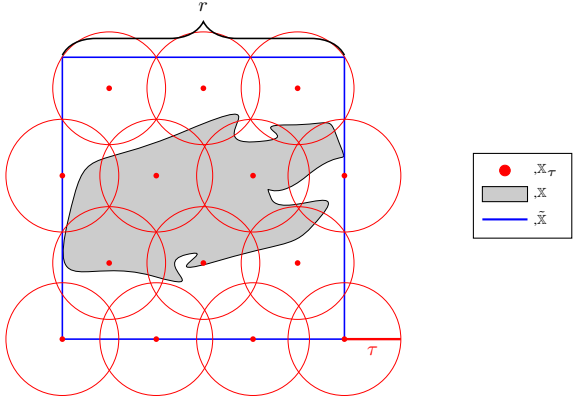

The virtual grid constant $\tau$ used in \eqref{eq:beta} and \eqref{eq:gamma}
balances the effect of the state space discretization and the inherent uncertainty 
measured by the posterior standard deviation $\sigma(\cdot)$. Therefore,
$\gamma(\tau)$ can be made arbitrarily small by choosing a sufficiently fine
virtual grid. This in turn increases $\beta_{\mathbb{X}}(\tau)$ and thus the 
effect of the posterior standard deviation $\sigma(\cdot)$ on the bound. 
However, $\beta_{\mathbb{X}}(\tau)$ depends merely logarithmically on $\tau$ such that
even poor Lipschitz constants $L_{\mu}$, $L_f$ and moduli of continuity $\omega_{\sigma}(\cdot)$ can
be easily compensated by small virtual grid constants $\tau$. 

\begin{remark}
	Since the standard deviation $\sigma(\cdot)$ varies within the state space $\mathbb{X}$,
	an optimal virtual grid constant $\tau$, which minimizes the expression 
	$\sqrt{\beta_{\mathbb{X}}(\tau)}\sigma(\bm{x})+\gamma(\tau)$ for all $\bm{x}\in\mathbb{X}$, does 
	not exist in general. While simple approaches such as choosing $\tau$ such that 
	$\gamma(\tau)$ is negligible for all $\bm{x}\in\mathbb{X}$ provide satisfying 
	results in our simulations, more complex approaches remain open research questions.
\end{remark}

It is important to note that most of the parameters in \cref{th:errbound_with}
do not require a difficult analysis such that the bound \eqref{eq:errorbound} 
can be directly evaluated. While the computation of the exact covering number 
$M(\tau,\mathbb{X})$ is a difficult problem for general sets $\mathbb{X}$,
it can be easily upper bounded as illustrated in \cref{fig:cover}. For this reason, 
we overapproximate the set 
$\mathbb{X}$ through a $d$-dimensional hypercube $\tilde{\mathbb{X}}$ with edge length $r$. 
Then, the covering number of $\tilde{\mathbb{X}}$ is bounded by \cite{Shalev-Shwartz2013} 
\begin{align}
M(\tau,\tilde{\mathbb{X}})\leq \left(\frac{r\sqrt{d}}{2\tau}\right)^d,
\end{align}
which is by construction also a bound for the covering number of $\mathbb{X}$, 
i.e., 
\begin{align}
M(\tau,\mathbb{X})\leq \left(\frac{r\sqrt{d}}{2\tau}\right)^d.
\end{align}

The Lipschitz constant $L_{\mu}$ of the posterior mean in \eqref{eq:gamma} can 
be straightforwardly bounded when the prior Gaussian process has a Lipschitz continuous 
kernel, as shown in the following lemma.
\begin{lemma}\label{Lipschitz mean}
Consider a zero mean prior Gaussian process defined through the $L_k$-Lipschitz 
	kernel $k(\cdot,\cdot)$. Then, its 
	posterior mean $\mu(\cdot)$  
	is continuous with Lipschitz constant\looseness=-1
	\begin{align}
	L_{\mu}&\leq L_k\sqrt{N} 
	\left\| (\bm{K}+\sigma_{\mathrm{on}}^2\bm{I}_N)^{-1}\bm{y} \right\|.
	\end{align}
\end{lemma}
\begin{IEEEproof}
The norm of the difference between 
	the posterior mean $\mu(\bm{x})$ evaluated at two different points is given by
	\begin{align}
	\|\mu(\bm{x})-\mu(\bm{x}')\|&=
	\left\| \left(\bm{k}(\bm{x})-\bm{k}(\bm{x}')\right)
	\bm{\alpha}\right\|,
	\end{align}
	with 
	\begin{align}
	\bm{\alpha}=(\bm{K}+\sigma_{\mathrm{on}}^2\bm{I}_N)^{-1}\bm{y}.
	\end{align}
	Due to the Cauchy-Schwarz inequality and the Lipschitz continuity of the 
	kernel we obtain
	\begin{align}
	\|\mu(\bm{x})-\mu(\bm{x}')\|&\leq L_k\sqrt{N} \left\| \bm{\alpha} 
	\right\|\|\bm{x}-\bm{x}'\|,
	\end{align}
	which proves Lipschitz continuity of the mean $\mu(\bm{x})$.
\end{IEEEproof}

Moreover, the assumption of a Lipschitz continuous kernel also suffices to compute the modulus of 
continuity $\omega_{\sigma}(\cdot)$ for the posterior standard deviation in \eqref{eq:gamma}, 
as shown in the following lemma.\looseness=-1
\begin{lemma}\label{Lipschitz variance}
Consider a zero mean prior Gaussian process defined through the $L_k$-Lipschitz 
	kernel $k(\cdot,\cdot)$. Then, its 
	posterior standard deviation $\sigma^2(\cdot)$  
	is continuous with modulus of continuity\looseness=-1
	\begin{align}
	\omega_{\sigma}(\tau)&\leq \sqrt{2L_k\tau}.
	\end{align}
\end{lemma}
\begin{IEEEproof}
The difference between two different evaluations of the posterior standard deviation is bounded by
\begin{align}
|\sigma(\bm{x})-\sigma(\bm{x}')|\leq d_k(\bm{x},\bm{x}')
\end{align}
as shown in \cite{Curi2020}, where the kernel metric is defined as 
\begin{align}\label{eq:kernMet}
d_k(\bm{x},\bm{x}')=\sqrt{k(\bm{x},\bm{x})+k(\bm{x}',\bm{x}')-2k(\bm{x},\bm{x}')}.
\end{align}
Due to Lipschitz continuity of the kernel, we have
\begin{align}
d_k(\bm{x},\bm{x}')\leq \sqrt{2L_k\|\bm{x}-\bm{x}'\|},
\end{align}
which concludes the proof.
\end{IEEEproof}

For the special case of stationary kernels \mbox{$k(\bm{x},\bm{x}')=k(\bm{x}\!-\!\bm{x}')$}, the convergence 
rate of the modulus of continuity $\omega_{\sigma}(\cdot)$ can even be improved, as shown in the
following.
\begin{corollary}\label{cor:Lipschitz var stat}
Consider a zero mean prior Gaussian process defined through the stationary, $L_k$-Lipschitz 
	kernel $k(\cdot,\cdot)$. Then, its 
	posterior standard deviation $\sigma(\cdot)$  
	is continuous with modulus of continuity $\omega_{\sigma}(\tau)=L_{\sigma}\tau$, where \looseness=-1
	\begin{align}\label{eq:Lsigma}
	L_{\sigma}\\
	&= \sup\limits_{\bm{x}-\bm{x}'\in\mathbb{X}}\sqrt{\frac{1}{2k(\bm{0})\!-\!2k(\bm{x}\!-\!\bm{x}')}}\left\|\nabla k(\bm{x}\!-\!\bm{x}')\right\|.
	\end{align}
\end{corollary}
\begin{IEEEproof}
For stationary kernels, we can express the kernel metric as
\begin{align}\label{eq:kernMet_stat}
d_k(\bm{x},\bm{x}')=d_k(\bm{x}-\bm{x}')=\sqrt{2k(\bm{0})-2k(\bm{x}-\bm{x}')}.
\end{align}
The simplified kernel 
metric is only a function of $\bm{x}-\bm{x}'$, such that the supremum of the norm of the 
derivative of $d_k(\cdot,\cdot)$ with respect to $\bm{x}-\bm{x}'$ is the Lipschitz constant of $\sigma(\cdot)$.
This derivative directly follows from the chain rule of differentation as
\begin{align}
\nabla d_k(\bm{x}-\bm{x}') = \sqrt{\frac{1}{2k(\bm{0})-2k(\bm{x}-\bm{x}')}}\nabla k(\bm{x}-\bm{x}'),
\end{align}
which concludes the proof.
\end{IEEEproof}
While computing the Lipschitz constant $L_{\sigma}$ requires the computation of a supremum in general, 
this optimization problem can be straightforwardly solved analytically for specific kernel choices, e.g., 
squared exponential kernels \cite{Lederer2021b}. Thereby, it allows the efficient computation of a tight
modulus of continuity.

The remaining open parameter in \eqref{eq:gamma} is the Lipschitz constant $L_f$ of the unknown
function $f(\cdot)$. In many applications, in particular in control, rough knowledge of the
unknown function is known in advance, which can allow to specify $L_f$. Even if this constant
is a rather poor estimate of the true Lipschitz constant, conservative estimates are not a crucial
issue as discussed after \cref{th:errbound_with}. If no such knowledge of the unknown function $f(\cdot)$
is available, the prior Gaussian process distribution can be employed to derive a probabilistic 
Lipschitz constant as shown in the following section.

\subsection{Probabilistic Lipschitz Constants for Gaussian Processes}
\label{subsec:prob Lipschitz}

In order to derive a probabilistic Lipschitz constant $L_f$ of the unknown function $f(\cdot)$ from
the prior Gaussian process distribution, we exploit the fact that the derivative of a Gaussian process
is again a Gaussian process. Therefore, Lipschitz constants can be obtained by adapting results from
the well-studied theory of suprema of Gaussian processes. This yields the following 
lemma, which is based on the metric entropy criterion~\cite{Dudley1967}.
\begin{lemma}
	\label{lem:expsup}
	Consider a Gaussian process with a continuously differentiable covariance 
	function $k(\cdot,\cdot)$ and let $L_k$ denote its Lipschitz constant on 
	the compact set $\mathbb{X}$ which is included in a cube with edge length $r$. 
	Then, the expected supremum of a sample function $f(\cdot)$ of this
	Gaussian process satisfies
	\begin{align}\label{eq:Esup}
	E\left[\sup\limits_{\bm{x}\in \mathbb{X}}f(\bm{x})\!\right]\leq 
	12\sqrt{6d}\max\!\left\{\!\max\limits_{\bm{x}\in\mathbb{X}}\!\sqrt{k(\bm{x},\bm{x})},\sqrt{rL_k}\!\right\}.
	\end{align}
\end{lemma}
\begin{IEEEproof}
	We prove this lemma by making use of the metric entropy criterion for the 
	sample continuity of Gaussian processes \cite{Dudley1967}.
	This criterion allows to bound the expected supremum of a sample function $f(\cdot)$
	by
	\begin{equation}
	\mathrm{E}\left[ \sup\limits_{\bm{x}\in\mathbb{X}}f(\bm{x}) \right]\leq \int\limits_0^{\max\limits_{\bm{x}\in\mathbb{X}}\sqrt{k(\bm{x},\bm{x})}}
	\sqrt{\log(N_k(\varrho,\mathbb{X}))}\mathrm{d}\varrho,
	\label{eq:metEntropy}
	\end{equation}
	where $N_k(\varrho,\mathbb{X})$ is the $\varrho$-packing number of $\mathbb{X}$ 
	with respect to the kernel metric \eqref{eq:kernMet}.
	Instead of bounding the $\varrho$-packing number, we bound the $\varrho/2$-covering 
	number, which is known to be an upper bound of the packing number. The covering number can be easily bounded 
	by transforming the problem of covering $\mathbb{X}$ with respect to the metric 
	$d_k(\cdot,\cdot)$ into a coverage problem in the original metric of $\mathbb{X}$. For 
	this reason, define
	\begin{align}
	\psi(\varrho')=\sup\limits_{\subalign{\bm{x},\bm{x}'&\in \mathbb{X}\\ 
			\|\bm{x}-\bm{x}'&\|_{\infty}\leq \varrho'}} d_k(\bm{x},\bm{x}'),
	\end{align}
	which is continuous due to the continuity of the covariance kernel $k(\cdot,\cdot)$. 
	Consider the inverse function
	\begin{align}
	\psi^{-1}(\varrho)=\inf\left\{\varrho'>0:~\psi(\varrho')>\varrho\right\}.
	\end{align}
	Continuity of $\psi(\cdot)$ implies $\varrho=\psi(\psi^{-1}(\varrho))$. In particular, 
	this means that we can guarantee $d_k(\bm{x},\bm{x}')\leq \frac{\varrho}{2}$ if 
	\mbox{$\|\bm{x}-\bm{x}'\|\leq \psi^{-1}(\frac{\varrho}{2})$}. Due to this relationship 
	it is sufficient to construct a uniform grid with grid constant $2\psi^{-1}(\frac{\varrho}{2})$ 
	in order to obtain a $\varrho/2$-covering net of $\mathbb{X}$. Furthermore, the cardinality 
	of this grid is an upper bound for the $\varrho/2$-covering number, such that we obtain
	\begin{align}
	N_k(\varrho,\mathbb{X})\leq 
	\left\lceil \frac{r}{2\psi^{-1}(\frac{\varrho}{2})} \right\rceil^{d}.
	\end{align}
	Due to the Lipschitz continuity of the covariance function, we can bound 
	$\psi(\cdot)$ by $\psi(\varrho')\leq \sqrt{2L_k\varrho'}.$
	Hence, the inverse function satisfies
	\begin{align}
	\psi^{-1}\left(\frac{\varrho}{2}\right)\geq 
	\left(\frac{\varrho}{2\sqrt{2L_k}}\right)^2
	\end{align}
	and consequently
	\begin{align}
	N_k(\varrho,\mathbb{X})\leq \left(1+\frac{4rL_k}{\varrho^2}\right)^{d}
	\end{align}
	holds, where the ceil operator is resolved through the addition of $1$.
	Substituting this expression in the metric entropy bound 
	\eqref{eq:metEntropy} yields
	\begin{align}
	\!E\!\left[\sup\limits_{\bm{x}\in \mathbb{X}}f(\bm{x})\!\right]\leq 12\sqrt{d}
	\!\!\int\limits_0^{\max\limits_{\bm{x}\in\mathbb{X}}\sqrt{k(\bm{x},\bm{x})}}
	\!\!\sqrt{\log\left(\!1\!+\!\frac{4rL_k}{\varrho^2}\!\right)}\mathrm{d}\varrho.\!
	\end{align}
	As shown in \cite{Grunewalder2010} this integral can be bounded by $\sqrt{6}\max\{\max_{\bm{x}\in\mathbb{X}}\sqrt{k(\bm{x},\bm{x})}, \sqrt{rL_k}\}$, which concludes the proof.
\end{IEEEproof}

While \cref{lem:expsup} provides a bound merely for the expected supremum of 
a sample function, a high probability bound for the supremum can be obtained using
the Borell-TIS inequality \cite{Talagrand1994}. This is shown in the following 
result.
\begin{lemma}
	\label{lem:supbound}
	Consider a Gaussian process with a continuously differentiable covariance 
	function $k(\cdot,\cdot)$. 
	Then, with probability of at least $1-\delta_L$ the supremum of a sample
	function $f(\cdot)$ of this Gaussian process is bounded by 
	\begin{align}\label{eq:fsup}
	f_{\mathrm{sup}}(\delta_L,k(\cdot,\cdot),r)=&
	\sqrt{2\log\left( \frac{1}{\delta_L} \right)}\max\limits_{\bm{x}\in\mathbb{X}}
	\sqrt{k(\bm{x},\bm{x})}\\
	&+12\sqrt{6d}
	\max\!\left\{\!\max\limits_{\bm{x}\in\mathbb{X}}\!\sqrt{k(\bm{x},\bm{x})},
	\sqrt{rL_{k}}\!\right\}.\nonumber
	\end{align}
\end{lemma}
\begin{IEEEproof}
	We prove this lemma by exploiting the wide theory of concentration inequalities 
	to derive a bound for the supremum of the sample function $f(\bm{x})$. We apply the 
	Borell-TIS inequality \cite{Talagrand1994}, which ensures for arbitrary $c\in\mathbb{R}_{0,+}$ that
	\begin{align}
	P\!\left( \sup\limits_{\bm{x}\in\mathbb{X}}f(\bm{x})\!-\!
	E\left[ \sup\limits_{\bm{x}\in\mathbb{X}}f(\bm{x}) \!\right] \!\geq\! 
	c \!\right)\leq \exp\left(\!\! -\frac{c^2}{2\max\limits_{\bm{x}\in\mathbb{X}}
	k(\bm{x},\bm{x})} \!\right)\!.
	\label{eq:Tal}
	\end{align}
	Due to \cref{lem:expsup} we can directly bound $E[\sup_{\bm{x}\in \mathbb{X}}f(\bm{x})]$. Therefore, 
	the lemma follows from 
	substituting \eqref{eq:Esup} in \eqref{eq:Tal} and choosing 
	$c=\sqrt{2\log\left( \nicefrac{1}{\delta_L} \right)}\max_{\bm{x}\in\mathbb{X}}\sqrt{k(\bm{x},\bm{x})}$.
\end{IEEEproof}

Since the derivatives of sample functions from Gaussian processes with sufficiently smooth kernels are the 
sample functions of the derivative Gaussian processes \cite{Ghosal2006}, \cref{lem:supbound} directly 
allows to compute a high probability Lipschitz constant for the unknown function $f(\cdot)$ from the prior
Gaussian process distribution. This is summarized in the following Theorem. 
\begin{theorem}
	\label{th:errbound_without}
	Consider a zero mean Gaussian process defined through the
	covariance kernel $k(\cdot,\cdot)$ 
	with continuous partial derivatives up to the fourth order
	and partial derivative kernels 
	\begin{align}
	k^{\partial i}(\bm{x},\bm{x}')&=\frac{\partial^2}{\partial x_i\partial x_i'}
	k(\bm{x},\bm{x}')\quad \forall i=1,\ldots, d.
	\end{align}
	Then, a sample function $f(\cdot)$ of the Gaussian process is almost 
	surely continuous on $\mathbb{X}$ and with probability of at least $1-\delta_L$, 
	\begin{align}
	L_f\leq\hat{L}_f=\left\|\begin{bmatrix}
	f_{\mathrm{sup}}(\nicefrac{\delta_L}{2d},k^{\partial 1}(\cdot,\cdot),r)\\
	\vdots\\
	f_{\mathrm{sup}}(\nicefrac{\delta_L}{2d},k^{\partial d}(\cdot,\cdot),r)
	\end{bmatrix}\right\|
	\label{eq:Lfbound}
	\end{align}
	for $f_{\mathrm{sup}}(\cdot,\cdot,\cdot)$ defined in \eqref{eq:fsup}.
\end{theorem}
\begin{IEEEproof}
	Continuity of the sample function $f(\bm{x})$ follows directly from 
	\cite[Theorem 5]{Ghosal2006}. Furthermore, this theorem guarantees that the derivative 
	functions $\frac{\partial}{\partial x_i}f(\bm{x})$ are samples from derivative Gaussian 
	processes with covariance functions $k^{\partial i}(\bm{x},\bm{x}')$.
	Therefore, we can apply \cref{lem:supbound} to each of the derivative processes and 
	obtain with probability of at least $1-\frac{\delta_L}{d}$
	\begin{align}
	\sup\limits_{\bm{x}\in\mathbb{X}}\left|\frac{\partial}{\partial x_i}f(\bm{x})\right| 
	\leq f_{\mathrm{sup}}(\nicefrac{\delta_L}{2d},k^{\partial i}(\cdot,\cdot),r).
	\label{eq:partderbound}
	\end{align}
	Applying the union bound over all partial derivative 
	processes $i=1,\ldots,d$ finally yields the result.
\end{IEEEproof}

Since many practically employed kernels such as, e.g., the squared exponential, the Matern $\nicefrac{5}{2}$, 
satisfy the required smoothness assumption of \cref{th:errbound_without}, this assumption does not pose
a severe restriction. Therefore, this theorem allows to straightforwardly determine high probability Lipschitz constants
for the unknown function $f(\cdot)$, which can be directly used in \cref{th:errbound_with}, while barely requiring additional
assumptions.

\section{Data Dependency of Learning Error Bounds}
\label{sec:postVar}

In order to derive conditions for ensuring that the learning error bound in \cref{th:errbound_with} is
below a given threshold as described \cref{subsubsec:e2d}, we need to analyze its dependency on the training data density. For this purpose, we investigate the decay behavior of the probabilistic uniform error bound \eqref{eq:errorbound} depending on the decrease rate of the GP standard deviation in \cref{subsec:varCond}. A kernel-dependent measure of data density is proposed in \cref{subsec:varBound_gen} in order to bound the decrease rate of the GP standard deviation. Finally, it is shown in \cref{subsec:varbound_stat} how the kernel-dependent density measure can be bounded using straightforwardly computable Euclidean distances.

\subsection{Asymptotic Bounds for the Learning Error}\label{subsec:varCond}

Since the probabilistic uniform error bound \eqref{eq:errorbound} consists of two summands, a vanishing posterior standard deviation $\sigma(\bm{x})$ is not by itself sufficient to guarantee a decreasing value of~$\eta(\bm{x})$. Therefore, it is necessary to additionally vary the parameter~$\tau$, such that $\gamma(\tau)$ decreases with growing number of training samples $N$. Even though this leads to a growing value of $\beta_{\mathbb{X}}(\tau)$, it ensures an asymptotically vanishing learning error bound in the limits $N\rightarrow \infty$ and $\sigma(\bm{x})\rightarrow 0$ as shown in the following theorem.
\begin{theorem}
	\label{th:as_err}
	Consider a zero mean Gaussian process defined by the continuously differentiable 
	kernel $k(\cdot,\cdot)$. Let $f:\mathbb{X}\rightarrow \mathbb{R}$ be a continuous unknown function with Lipschitz constant $L_f$ on the compact domain $\mathbb{X}$ which satisfies \cref{ass:samplefun}. 
	Then, for $\tau\in\mathcal{O}(\nicefrac{1}{N})$, the learning error asymptotically behaves as 
	\begin{align}
		\eta(\bm{x})\in\mathcal{O}\!\left(\!\!\sqrt{\log\!\left(\!\frac{N}{\delta}\right)}\sigma(\bm{x})\!+\!\frac{1}{N}\!\!\right).
	\end{align}
\end{theorem}
\begin{IEEEproof}
	Due to Theorem~\ref{th:errbound_with} with suitable value of $\beta_{\mathbb{X}}(\tau)$ it holds that
	\begin{align}
	\!\sup\limits_{\bm{x}\in \mathbb{X}}|f(\bm{x})\!-\!\mu(\bm{x})|\!\leq\!
	\sqrt{\beta_{\mathbb{X}}(\tau)}\sigma(\bm{x})\!+\!\gamma(\tau)
	\label{eq:un_err}
	\end{align}
	with probability of at least $1-\delta/2$ for $\delta\in(0,1)$. A trivial 
	bound for the covering number can be obtained by considering a uniform grid over the 
	cube containing $\mathbb{X}$. This approach leads to
	\begin{align}
	M(\tau,\mathbb{X})\leq \left(\frac{r\sqrt{d}}{2\tau}\right)^{d}.
	\end{align}
	Therefore,
	we have
	\begin{align}
	\beta_{\mathbb{X}}(\tau)\leq 2d\log\left(\frac{r\sqrt{d}}{2\tau}\right)-2\log(\delta).
	\label{eq:beta(tau)}
	\end{align}
	In order to derive a bound for $\gamma(\tau)$, we employ the bounds for the Lipschitz constants and
	modulus of continuity. 
	The Lipschitz constant $L_{\mu}$ in \eqref{eq:gamma} is bounded by 
	\begin{align}
	L_{\mu}&\leq L_k\sqrt{N} 
	\left\| (\bm{K}+\sigma_{\mathrm{on}}^2\bm{I}_N)^{-1}\bm{y} \right\|
	\end{align}
	due to \cref{Lipschitz mean}. Since the Gram matrix $\bm{K}$ 
	is positive semidefinite and $f(\cdot)$ is bounded by some $\bar{f}$ due to Lipschitz continuity and a compact domain $\mathbb{X}$, we can 
	bound $\left\| (\bm{K}+\sigma_{\mathrm{on}}^2\bm{I}_N)^{-1}\bm{y} \right\|$ by
	\begin{align}
	\left\| (\bm{K}+\sigma_{\mathrm{on}}^2\bm{I}_N)^{-1}\bm{y} \right\|&
	\leq\frac{\|\bm{y}\|}
	{\lambda_{\min}(\bm{K}+\sigma_{\mathrm{on}}^2\bm{I}_N)}\nonumber\\
	&\leq \frac{\sqrt{N}\bar{f}
		+\|\bm{\epsilon}\|}{\sigma_{\mathrm{on}}^2},
	\end{align}
	where $\bm{\epsilon}$ is a vector of $N$ i.i.d. zero mean Gaussian random variables with
	variance $\sigma_{\mathrm{on}}^2$. Therefore, it follows that 
	$\frac{\|\bm{\epsilon}\|^2}{\sigma_{\mathrm{on}}^2}\sim\chi_N^2$. Due to 
	\cite{Laurent2000}, with probability of at least $1-\exp(-\log(\nicefrac{2}{\delta}))$ we have 
	\begin{align}\label{eq:noisebound}
	\|\bm{\epsilon}\|^2\leq \left(2\sqrt{N\log\left(\frac{2}{\delta}\right)}+2\log\left(\frac{2}{\delta}\right)+N\right)\sigma_{\mathrm{on}}^2.
	\end{align}
	Hence, the Lipschitz constant of the 
	posterior mean function $\mu(\cdot)$ satisfies with probability of at least 
	$1-\nicefrac{\delta}{2}$
	\begin{align}\label{eq:Lmu_bound}
	L_{\mu}\leq L_k\frac{N\bar{f}+\sqrt{N\left(2\sqrt{N\log\left(\frac{2}{\delta}\right)}+2\log\left(\frac{2}{\delta}\right)+N\right)}\sigma_{\mathrm{on}}}
	{\sigma_{\mathrm{on}}^2}.
	\end{align}
	It can clearly be seen that the fastest growing term is increasing linearly, such that it holds that $L_{\mu}\in\mathcal{O}(N)$ with 
	probability of at least $1-\nicefrac{\delta}{2}$. 
	The modulus of continuity in \eqref{eq:gamma} can be bounded by
	\begin{align}\label{eq:omega_bound}
		\omega_{\sigma}(\tau)\leq \sqrt{2L_k\tau}
	\end{align}
	due to \cref{Lipschitz variance}. 
	Since the unknown function $f(\cdot)$ is assumed
	to admit a Lipschitz 
	constant $L_f$, we obtain
	\begin{align}\label{eq:gamma_bound}
	\gamma(\tau)\leq&L_k\tau\frac{N\bar{f}+\sqrt{N\left(2\sqrt{N\log\left(\frac{2}{\delta}\right)}+2\log\left(\frac{2}{\delta}\right)+N\right)}\sigma_{\mathrm{on}}}
	{\sigma_{\mathrm{on}}^2}\nonumber\\
	&+\sqrt{2\beta_{\mathbb{X}}(\tau)L_k\tau}
	+L_f\tau.
	\end{align}
	with probability of at least $1-\nicefrac{\delta}{2}$ by substituting \eqref{eq:Lmu_bound} and \eqref{eq:omega_bound} into \eqref{eq:gamma}. In order to admit asymptotically vanishing 
	error bounds, \eqref{eq:gamma_bound} must converge to $0$ for $N\rightarrow\infty$, which is only ensured if $\tau$ decreases faster 
	than $\mathcal{O}(1/N)$. Therefore, set $\tau\in\mathcal{O}(1/N)$ in order 
	to guarantee
	\begin{align}
	\gamma_N(\tau)\in\mathcal{O}\left( \frac{1}{N} \right).
	\end{align}
	However, this choice of $\tau$ implies that $\beta_{\mathbb{X}}(\tau)\in\mathcal{O}(\log(\frac{N}{\delta}))$ 
	due to \eqref{eq:beta(tau)}. Therefore, it directly follows that
	\begin{align}
	\!\!\sqrt{\beta_{\mathbb{X}}(\tau)}\sigma(\bm{x})\!+\!\gamma(\tau)\!\in\!\mathcal{O}\!\left(\!\!\sqrt{\log\!\left(\!\!\frac{N}{\delta}\!\right)}\sigma(\bm{x})\!+\!\frac{1}{N}\!\!\right)\!\!,\!\!
	\label{eq:ass_analysis}
	\end{align}
	which concludes the proof.
\end{IEEEproof}

Due to the linear dependency of the bound for the Lipschitz constant $L_{\mu}$ on the number
of training samples, the virtual grid constant must decay faster than $\mathcal{O}(\nicefrac{1}{N})$. 
This in turn leads to a logarithmic growth of $\beta_{\mathbb{X}}(\tau)$, which causes the $\sqrt{\log(N)}$ 
increase of the scaling factor of the posterior standard deviation $\sigma(\bm{x})$. Note that this is a 
common phenomenon in uniform error bounds for GP regression and can also be found in RKHS based approaches, where
similar bounds as \eqref{eq:noisebound} are used to bound the effect of the noise \cite{Srinivas2012,Chowdhury2017a}.

\subsection{Asymptotic Bounds for the Posterior Variance}
\label{subsec:varBound_gen}

In order to compensate the growth of the scaling factor in \cref{th:as_err}, a sufficiently 
fast decay of the standard deviation $\sigma(\bm{x})$ must be ensured. Therefore, we investigate
the behavior of the posterior variance $\sigma^2(\bm{x})$ depending on the training data density of an input data set
$\mathbb{D}^x=\{\bm{x}^{(i)}\}_{i=1}^{N}$. The starting point of this analysis
is the following lemma, which provides a straightforward upper bound for the 
posterior variance $\sigma^2(\bm{x})$.%

\begin{lemma}\label{lem:sigma_bound}
	Consider a GP trained using a 
	data set with input training samples~$\mathbb{D}^x$. Then, 
	the posterior variance is bounded by\looseness=-1
	\begin{align}\label{eq:var_bound_base}
		\sigma^2(\bm{x})&\leq 
		\frac{\sigma_{\mathrm{on}}^2k(\bm{x},\bm{x})+N\Delta k(\bm{x})}{N
			\max\limits_{\bm{x}'\in\mathbb{D}^x}
			k(\bm{x}',\bm{x}')+\sigma_{\mathrm{on}}^2},
	\end{align}
	where
	\begin{align}\label{eq:Deltak}
		\!\Delta k(\bm{x})= k(\bm{x},\bm{x})\max\limits_{\bm{x}'\in\mathbb{D}^x}
		k(\bm{x}',\bm{x}') \!-\! \min\limits_{\bm{x}'\in\mathbb{D}^x}
		k^2(\bm{x}',\bm{x}).\!
	\end{align}
\end{lemma}
\begin{IEEEproof}
	Since $\bm{K}+\sigma_{\mathrm{on}}^2\bm{I}_N$ is a 
	positive definite, quadratic matrix, it follows 
	that
	\begin{align}
	\sigma^2(\bm{x})&\leq k(\bm{x},\bm{x})-
	\frac{\left\|\bm{k}(\bm{x})\right\|^2}
	{\lambda_{\max}\left(\bm{K}\right)+\sigma_{\mathrm{on}}^2}.
	\end{align}
	Applying the Gershgorin theorem \cite{Gershgorin1931} 
	the maximal eigenvalue is bounded by
	\begin{align}
	\lambda_{\max}(\bm{K})\leq N
	\max\limits_{\bm{x}'\in\mathbb{D}^x}
	k(\bm{x}',\bm{x}').
	\end{align}
	Furthermore, due to the definition of 
	$\bm{k}(\bm{x})$ we have
	\begin{align}
	\|\bm{k}(\bm{x})\|^2\geq N 
	\min\limits_{\bm{x}'\in\mathbb{D}^x}
	k^2(\bm{x}',\bm{x}).
	\end{align}
	Therefore, $\sigma^2(\bm{x})$ can be bounded by
	\begin{align}
	\sigma^2(\bm{x})&\leq k(\bm{x},\bm{x})-
	\frac{N\min\limits_{\bm{x}'\in\mathbb{D}^x}
		k^2(\bm{x}',\bm{x})}{N
		\max\limits_{\bm{x}'\in\mathbb{D}^x}
		k(\bm{x}',\bm{x}')+\sigma_{\mathrm{on}}^2}.
	\label{eq:sigma_bound1}
	\end{align}
	Finally, the proof follows from the definition of $\Delta k(\bm{x})$.
\end{IEEEproof}
This theorem does not pose any restriction on the employed kernel, but strongly depends on
the particular choice of kernel. Therefore, it can be difficult to interpret. However, it can be significantly 
simplified for specific kernels, as shown in the following corollary for stationary 
covariance functions.
\begin{corollary}
	\label{cor:var_bound}
	Consider a GP with stationary kernel and input training samples~$\mathbb{D}^x$. Then, 
	the posterior variance is bounded by\looseness=-1
	\begin{align}
	\sigma^2(\bm{x})\leq k(0)-\frac{\min\limits_{\bm{x}'\in\mathbb{D}^x}k^2(\bm{x}-\bm{x}')}{k(0)
		+\frac{\sigma_{\mathrm{on}}^2}{N}}.
	\label{eq:isobound}
	\end{align}
\end{corollary}
\begin{IEEEproof}
	The proof follows directly from 
	\cref{lem:sigma_bound} and the fact that $\max_{\bm{x}'\in\mathbb{D}^x}k(\bm{x}',\bm{x}')= k(\bm{0})$
	since the kernel is stationary.
\end{IEEEproof}
In this special case of \cref{lem:sigma_bound}, which has been previously stated, 
e.g., in \cite{Shekhar2018}, the kernel induces a notion of proximity, where the absence 
of training inputs $\bm{x}'$ with $k(\bm{x}\!-\!\bm{x}')\approx 0$ leads to a large bound for the 
posterior variance $\sigma^2(\bm{x})$. Therefore, this corollary shows that it is 
desirable to have data close to the test point $\bm{x}$ as measured by $k(\cdot)$ for
stationary kernels. 

Since \cref{lem:sigma_bound} and \cref{cor:var_bound} still consider the full input data 
set $\mathbb{D}^x$, a single sample with $k(\bm{x}',\bm{x})\approx 0$ can practically lead to the 
trivial bound $\sigma^2(\bm{x})\lesssim k(\bm{x},\bm{x})$. This is clearly an undesired behavior for
a bound since it would imply that additional data can potentially increase the posterior variance bound.
In order to avoid this effect, we make use of an important property of 
Gaussian process posterior variances, which is the fact that $\sigma^2(\bm{x})$ is 
non-increasing with the number of training samples $N$ \cite{Vivarelli1998}. Therefore,
we can consider subsets of $\mathbb{D}^x$ to compute the posterior variance
bounds in \cref{lem:sigma_bound} and \cref{cor:var_bound}, which exclude these training samples with 
a negative effect on the bound.
Due to the importance of $\Delta k(\bm{x})$ for these bounds, we make use of the following subset
\begin{align}\label{eq:Bset}
	\!\!\mathbb{K}_{\rho'}\!(\bm{x})&\!=\!\{ \bm{x}'\!\!\in\!\mathbb{D}^x\!\!: k^2\!(\bm{x},\bm{x})\!\leq\! k^2\!(\bm{x}'\!,\bm{x}')\!\leq\! \frac{1}{\rho'}\!\!+\!k^2\!(\bm{x}'\!,\bm{x}) \}\!\!\!
\end{align}
for this purpose. It can be easily seen that considering only the subset $\mathbb{K}_{\rho'}\!(\bm{x})\subset\mathbb{D}^x$ in \eqref{eq:Deltak} ensures 
\begin{align}
	k(\bm{x},\bm{x})\max\limits_{\bm{x}'\in\mathbb{K}_{\rho'}\!(\bm{x})}
	k(\bm{x}',\bm{x}') \!-\! \min\limits_{\bm{x}'\mathbb{K}_{\rho'}\!(\bm{x})}
	k^2(\bm{x}',\bm{x})\leq \frac{1}{\rho'}.
\end{align}  
Since the consideration of a subset of $\mathbb{D}^x$ also reduces the number of considered training 
samples in \eqref{eq:var_bound_base}, we trade-off the size of $\mathbb{K}_{\rho'}\!(\bm{x})$ and the 
ensured value for $\Delta k(\bm{x})$ by defining $\rho'$ using the following optimization problem
\begin{align}
\rho(\bm{x})=&\max\limits_{\rho'\in\mathbb{R}_+} \rho'\\
&\text{such that } |\mathbb{K}_{\rho'}(\bm{x})|\geq \rho'\sigma_{\mathrm{on}}^2k(\bm{x},\bm{x}).\label{eq:kernel_neighborhood}
\end{align}
It can easily be seen that $\rho(\bm{x})$ is well-defined since the optimization problem is always 
feasible for $\rho'\rightarrow 0$. Moreover, it can be directly used as a measure of data density 
as shown in the following proposition.
\begin{proposition}\label{prop:rho_decay}
	Consider a zero mean Gaussian process defined by the
	kernel $k(\cdot,\cdot)$. If $k(\bm{x},\bm{x})\neq 0$, the posterior standard deviation at~$\bm{x}$ 
	satisfies 
	\begin{align}
		\sigma(\bm{x})\leq \sqrt{\frac{2}{\rho(\bm{x})k(\bm{x},\bm{x})}}
	\end{align}
	such that it behaves as $\sigma(\bm{x})\in\mathcal{O}( \nicefrac{1}{\sqrt{\rho(\bm{x})}} )$.
\end{proposition}
\begin{IEEEproof}
	By exploiting the fact that the posterior variance $\sigma^2(\bm{x})$ is non-increasing with the number of training samples $N$ \cite{Vivarelli1998} 
	and considering only samples inside the set 
	$\mathbb{K}_{\rho(\bm{x})}(\bm{x})$ for the computation of the 
	posterior standard deviation, we obtain\looseness=-1
	\begin{align}
		\!\sigma^2(\bm{x})&\leq 
		\frac{\sigma_{\mathrm{on}}^2k(\bm{x},\bm{x})+|\mathbb{K}_{\rho(\bm{x})}(\bm{x})| \Delta k(\bm{x})}{|\mathbb{K}_{\rho(\bm{x})}(\bm{x})|
			\max\limits_{\bm{x}'\in\mathbb{K}_{\rho(\bm{x})}(\bm{x})}
			k(\bm{x}',\bm{x}')\!+\!\sigma_{\mathrm{on}}^2}\!
	\end{align}
	due to \cref{lem:sigma_bound}. Since $\bm{x}'\in\mathbb{K}_{\rho(\bm{x})}(\bm{x})$ implies $k(\bm{x}',\bm{x}')\geq k(\bm{x},\bm{x})$, we can simplify this expression to
	\begin{align}
		\sigma^2(\bm{x})&\leq 
		\frac{\sigma_{\mathrm{on}}^2}{|\mathbb{K}_{\rho(\bm{x})}(\bm{x})|
			}+\frac{\Delta k(\bm{x})}{k(\bm{x},\bm{x})}.
	\end{align}
	Moreover, it can be straightforwardly checked that the restriction to $\mathbb{K}_{\rho(\bm{x})}(\bm{x})$
	implies $\Delta k(\bm{x})\leq \nicefrac{1}{\rho(\bm{x})}$, which yields
	\begin{align}\label{eq:var_bound_aux}
		\sigma^2(\bm{x})&\leq 
		\frac{\sigma_{\mathrm{on}}^2k(\bm{x},\bm{x})}{|\mathbb{K}_{\rho(\bm{x})}(\bm{x})|
			k(\bm{x},\bm{x})}+\frac{1}{\rho(\bm{x})k(\bm{x},\bm{x})}
	\end{align}
	Since $|\mathbb{K}_{\rho(\bm{x})}(\bm{x})|$ is lower bounded by $\rho(\bm{x})\sigma_{\mathrm{on}}^2k(\bm{x},\bm{x})$ by definition, we obtain
	\begin{align}
		\sigma^2(\bm{x})&\leq \frac{2}{\rho(\bm{x})k(\bm{x},\bm{x})},
	\end{align}
	which directly implies $\sigma(\bm{x})\in\mathcal{O}(\nicefrac{1}{\sqrt{\rho(\bm{x})}})$.
	concluding the proof.
\end{IEEEproof}
It can be clearly seen that $\rho(\bm{x})$ is a measure of data density which is highly specific for each particular GP and therefore is capable of reflecting the requirements on good data distributions posed by the employed kernel $k(\cdot,\cdot)$. Moreover, it immediately follows from \cref{th:as_err} that a sufficiently fast growth of $\rho(\bm{x})$, i.e., $\rho(\bm{x})\notin\mathcal{O}(\log(N))$, guarantees a vanishing error bound $|\mu(\bm{x})-f(\bm{x})|\rightarrow 0$. Therefore, $\rho(\cdot)$ satisfies the requirements posed on a suitable measure of data density in \cref{subsubsec:e2d}.

\subsection{Conditions for Specific Kernels}\label{subsec:varbound_stat}

The high flexibility of \cref{prop:rho_decay} allows its application to GPs with arbitrary kernels, but comes at the price of a difficult interpretability. However, when we fix a specific kernel, it is often possible to derive more accessible and intuitive subsets contained in $\mathbb{K}_{\rho'}(\bm{x})$, as shown in the following lemma for linear, squared exponential and Mat\'ern class kernels.
\begin{lemma}\label{lem:K_to_Euclid}
	Geometrically interpretable subsets of $\mathbb{K}_{\rho'}(\bm{x})$ defined in \eqref{eq:Bset} are given by
	\begin{enumerate}
		\item the set
		\begin{align}
			\mathbb{H}_{\rho'}^c(\bm{x})=\Big\{\!&\bm{x}'\!\in\!\mathbb{D}^x\!: \|\bm{x}'\|^2(\|\bm{x}'\|^2-c\|\bm{x}\|^2) \!\leq\! \frac{1}{\rho'}, \\
			&\|\bm{x}\|\!\leq\!\|\bm{x}'\|, \left|\bm{x}^T\bm{x}'\right|\!\geq\! c\|\bm{x}\| \|\bm{x}'\| \!\Big\}\subset \mathbb{K}_{\rho'}(\bm{x})\nonumber
		\end{align}
%
%
		for every $c\in(0,1)$;\looseness=-1
		\item the Euclidean ball 
		\begin{align}
			&\mathbb{B}_{\sqrt{\nicefrac{1}{2L_{\partial k}\sigma_f^2\rho'}}}(\bm{x})=\\
			&\qquad\quad\left\{\bm{x}'\!\in\!\mathbb{D}^x\!: \|\bm{x}\!-\!\bm{x}'\|\leq \sqrt{\frac{1}{2L_{\partial k}\sigma_f^2\rho'}}  \right\} \subset \mathbb{K}_{\rho'}(\bm{x})\nonumber
		\end{align}
		for isotropic SE or Mat\'ern kernels with $\nu\geq \nicefrac{3}{2}$ and $\sigma_f^2=k(\bm{x},\bm{x})$.
	\end{enumerate}
\end{lemma}
\begin{IEEEproof}
	Due to the definition of the linear kernel, we have the identity
	\begin{align}\label{eq:Delta k_lin_1}
		k^2(\bm{x}',\bm{x}')-k^2(\bm{x}',\bm{x})= \|\bm{x}'\|^4-(\bm{x}^T\bm{x}')^2.
	\end{align}
	For $\nicefrac{|\bm{x}^T\bm{x}'|}{(\|\bm{x}\|\|\bm{x}'\|)}\geq c$, we therefore obtain
	\begin{align}
		k^2(\bm{x}',\bm{x}')-k^2(\bm{x}',\bm{x})\leq \|\bm{x}'\|^2\left(\|\bm{x}'\|^2-c\|\bm{x}\|^2\right).
	\end{align}
	Finally, the first inequality in \eqref{eq:Bset} yields the requirement
	\begin{align}
	k^2(\bm{x},\bm{x})=\|\bm{x}\|^4\leq \|\bm{x}'\|^4= k^2(\bm{x}',\bm{x}'),
	\end{align}
	which concludes the first part of the proof.
	For the second part of the proof, we exploit the continuous differentiability of Mat\'ern kernels with $\nu\geq \nicefrac{3}{2}$ and squared exponential kernels together with the fact that their derivative at $\bm{r}=\bm{x}-\bm{x}'=\bm{0}$ is $0$. Therefore, we have
	\begin{align}
	k(\bm{x}-\bm{x}')\geq \sigma_f^2-L_{\partial k}\|\bm{x}-\bm{x}'\|^2.
	\end{align}
	where $L_{\partial k}\in\mathbb{R}_+$ is the Lipschitz constant of the kernel derivative. 
	Using this lower bound, we obtain
	\begin{align}
	\!k^2(\bm{0})\!-\!k^2(\bm{x}-\bm{x}')\!&\leq\! 2L_{\partial k} \sigma_f^2\|\bm{x}\!-\!\bm{x}'\|^2\!-\!L_{\partial k}^2\|\bm{x}\!-\!\bm{x}'\|^4,\!
	\end{align}
	which we can simplify to 
	\begin{align}\label{eq:rho_bound aux}
	k^2(\bm{0})-k^2(\bm{x}-\bm{x}')&\leq 2L_{\partial k} \sigma_f^2\|\bm{x}-\bm{x}'\|^2
	\end{align}
	due to non-negativity of the norm. Therefore, $\|\bm{x}-\bm{x}'\|^2\leq \nicefrac{\rho'}{2L_{\partial k} \sigma_f^2}$ implies $|k^2(\bm{x},\bm{x})-k^2(\bm{x},\bm{x}')|\leq \rho'$. Since $k(\bm{x},\bm{x})=k(\bm{x}',\bm{x}')$ for isotropic kernels, the first inequality is always satisfied, concluding the proof.

\end{IEEEproof}

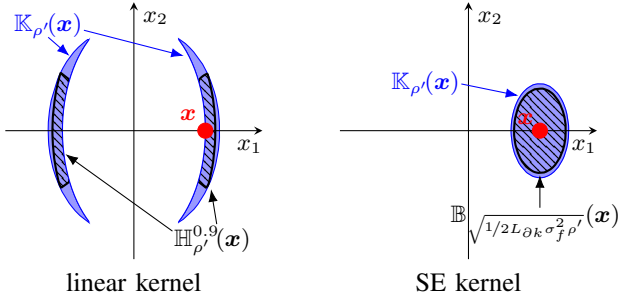
\begin{figure}
	\begin{minipage}{0.48\columnwidth}
		\centering 
		\def\file{files/lin_kern_vis.txt}	\tikzsetnextfilename{lin_kern_vis}
		\begin{tikzpicture}		
		\begin{axis}[inner axis line style={>={Latex[round]}},
		axis lines=center,yticklabels={,,},xticklabels={,,},ticks=none,
		grid=none,enlargelimits=false, axis on top,
		xlabel={$x_1$}, ylabel={$x_2$}, 
		xmin=-0.9, xmax = 0.9, ymin = -0.55, ymax = 0.55,	
		height=5cm, width=5cm, 
		x label style={at={(axis description cs:0.95,0.5)},anchor=north}
		]
		
		\addplot[blue, forget plot, name path = A] table[x = line3_1, y = line3_2]{\file};
		\addplot[blue, forget plot, name path = B] table[x = line4_1, y = line4_2]{\file};
		\addplot[blue, forget plot,fill=black, name path = C] table[x = line5_1, y = line5_2]{\file};
		\tikzfillbetween[of=A and C]{blue, opacity=0.4};
		\tikzfillbetween[of=B and C]{blue, opacity=0.4};
		\addplot[black,thick, forget plot,pattern=north west lines] table[x = line1_1, y = line1_2]{\file};
		\addplot[black,thick, forget plot,pattern=north west lines] table[x = line2_1, y = line2_2]{\file};

		\addplot[blue, forget plot,fill=white] table[x = line5_1, y = line5_2]{\file};
		\addplot[white, forget plot, fill=white] table[x = line6_1, y = line6_2]{\file};
		
		\pgfplotsset{
			after end axis/.code={
			\draw[red,fill=red] (axis cs:0.5,0) circle (0.1cm);
			
			\node at (axis cs:-0.6,0.45 ) {\textcolor{blue}{\small $\mathbb{K}_{\rho'}\!(\bm{x})$}};
			\node at (axis cs:0.38,0.07 ) {\textcolor{red}{\small $\bm{x}$}};
			\node at (axis cs:0.55,-0.47 ) {\textcolor{black}{\small $\mathbb{H}_{\rho'}^{0.9}\!(\bm{x})$}};
			
			\draw[-{Latex}, blue] (axis cs: -0.6,0.40) -- (axis cs:-0.45,0.32); 
			\draw[-{Latex}, blue] (axis cs: -0.35,0.43) -- (axis cs:0.35,0.32); 
				
			\draw[-{Latex}, black] (axis cs: 0.58,-0.4) -- (axis cs:0.54,-0.25); 
			\draw[-{Latex}, black] (axis cs: 0.27,-0.43) -- (axis cs:-0.48,-0.1); 
			
			\node at (axis cs:0.0,-0.65 ) {linear kernel};
			}
		}
		\end{axis}
		\end{tikzpicture}
	\end{minipage}\hfill
	\begin{minipage}{0.48\columnwidth}
		\centering 
		\def\file{files/SE_kern_vis.txt}	\tikzsetnextfilename{SE_kern_vis}
		\begin{tikzpicture}		
		\begin{axis}[inner axis line style={>={Latex[round]}},
		axis lines=center,yticklabels={,,},xticklabels={,,},ticks=none,
		grid=none,enlargelimits=false, axis on top,
		xlabel={$x_1$}, ylabel={$x_2$}, 
		xmin=-0.9, xmax = 0.9, ymin = -0.55, ymax = 0.55,	
		height=5cm, width=5cm, 
		x label style={at={(axis description cs:0.95,0.5)},anchor=north}
		 ]
		
		\addplot[blue, forget plot, fill=blue, opacity=0.4] table[x = line1_1, y = line1_2]{\file};
		\addplot[blue, forget plot] table[x = line1_1, y = line1_2]{\file};

		\addplot[black,thick, forget plot,pattern=north west lines] table[x = line2_1, y = line2_2]{\file};

		\pgfplotsset{
			after end axis/.code={
				\draw[red,fill=red] (axis cs:0.5,0) circle (0.1cm);
				
				\node at (axis cs:-0.3,0.2 ) {\textcolor{blue}{\small $\mathbb{K}_{\rho'}\!(\bm{x})$}};
				\node at (axis cs:0.4,0.05 ) {\textcolor{red}{\small $\bm{x}$}};
				\node at (axis cs:0.47,-0.4 ) {\textcolor{black}{\small $\mathbb{B}_{\sqrt{\nicefrac{1}{2L_{\partial k}\sigma_f^2\rho'}}}(\bm{x})$}};
				
				\draw[-{Latex}, blue] (axis cs: -0.05,0.2) -- (axis cs:0.35,0.15); 
				
				\draw[-{Latex}, black] (axis cs: 0.5,-0.33) -- (axis cs:0.5,-0.19); 
				
				\node at (axis cs:0.0,-0.65 ) {SE kernel};
			}
		}
		\end{axis}
		\end{tikzpicture}
	\end{minipage}
	\caption{Illustration of the set $\mathbb{K}_{\rho'}(\bm{x})$ and geometrically simple subsets for a linear and a SE kernel.}
	\label{fig:innersets}
\end{figure}
This lemma illustrates the flexibility of quantifying the data density using $\mathbb{K}_{\rho'}(\bm{x})$. While this set can
be innerapproximated by a ball for Mat\'ern and SE kernels as illustrated in \cref{fig:innersets}, it looks more like segments of a sphere for linear kernels. Since we can easily determine the volume of such simple geometrical structures, \cref{lem:K_to_Euclid} enables the derivation of a straightforward relationship between the sampling distributions and data density $\rho(\bm{x})$. For example, when training samples in $\mathbb{D}^{x}$ are generated by drawing from a uniform distribution, the number of points in a Euclidean ball is proportional to the volume of the ball, i.e., $\mathbb{B}_{\rho'}(\bm{x})\propto \nicefrac{N}{\rho'^d}$. Therefore, it follows from \eqref{eq:kernel_neighborhood} that $\rho(\bm{x})\in\mathcal{O}(N^{\nicefrac{1}{d+1}})$ for SE or Mat\'ern kernels with uniformly drawn input training samples. This in turn implies that  $\sigma(\bm{x})\in\mathcal{O}(\nicefrac{1}{N^{\nicefrac{1}{2d+2}}})$ due to \cref{prop:rho_decay} and consequently 
\begin{align}
	|\mu(\bm{x})-f(\bm{x})|\in\mathcal{O}\left(\frac{\log(N)}{N^{\nicefrac{1}{2d+2}}}\right)
\end{align}
due to \cref{th:as_err}. This demonstrates the flexibility and effectiveness of the derived formalism for bounding the asymptotic decay of the prediction error $|\mu(\bm{x})-f(\bm{x})|$ presented in this section.\looseness=-1

\section{Safety Guarantees for Control of Unknown Dynamical Systems}
\label{sec:safety}
We employ the theoretical results for GP error bounds introduced in the previous sections 
to develop an iterative approach for ensuring arbitrary tracking accuracy with the considered 
control law \eqref{eq:ctrl}. For this purpose, we derive a time-varying tracking 
error bound in \cref{subsec:stab} which depends explicitly on the uniform GP error bound along 
the reference trajectory. This result allows us to analyze the asymptotic decay of the tracking
error bound depending on the training data density measured by $\rho(\bm{x})$ in \cref{subsec:data2tracke}.
Finally, we employ the obtained insight to develop an episodic approach for ensuring arbitrary
tracking accuracy in \cref{subsec:epi_data}.

\subsection{Probabilistic Tracking Error Bound}
\label{subsec:stab}

Since \cref{as:stab} ensures distinct eigenvalues of the matrix $\bm{A}_{\bm{\theta}}$ defining the closed-loop behavior of the dynamics \eqref{eq:err_dyn} of the tracking error $\bm{e}=\bm{x}-\bm{x}_{\mathrm{ref}}$, we can compute the eigendecomposition $\bm{A}_{\bm{\theta}}=\bm{U}\bm{\Lambda}\bm{U}^{-1}$, where $\bm{\Lambda}$ is a diagonal matrix consisting of the eigenvalues of $\bm{A}_{\bm{\theta}}$. This allows the derivation of a dynamic bound for 
the tracking error $\bm{e}$ inspired by the comparison principle \cite{Khalil2002}, as shown in the following theorem.\looseness=-1

\begin{theorem}\label{th:tracking_time_varying}
	Consider a linear system \eqref{eq:sys} satisfying \cref{as:stab}, which is perturbed by 
	a $L_f$-Lipschitz nonlinearity $f(\cdot)$ satisfying \cref{ass:samplefun}. Assume that a zero
	mean Gaussian process with $L_k$-Lipschitz stationary kernel is used to learn a model $\hat{f}(\cdot)=\mu(\cdot)$ of $f(\cdot)$,
	such that a controller~\eqref{eq:ctrl} is used to track the bounded 
	reference $\bm{x}_{\mathrm{ref}}$. Then, 
	the tracking error is bounded by 
	\begin{align}\label{eq:ebound}
		\|\bm{x}(t)-\bm{x}_{\mathrm{ref}}(t)\|\leq \upsilon(t)
	\end{align}
	with probability of at least $1-\delta$, where $\upsilon(t)$ is the solution of the linear dynamical system 
	\begin{align}\label{eq:xi_diff}
		\dot{\upsilon}=\left(\lambda_{\max}(\bm{A}_{\bm{\theta}})+L_{\sigma}\zeta\sqrt{\beta_{\mathbb{X}}(\tau)}\right)\upsilon + \zeta \eta(\bm{x}_{\mathrm{ref}})
	\end{align}
	with initial condition $\upsilon(0)=\|\bm{U}\|\|\bm{U}\bm{e}(0)\|$ and constant  
	$\zeta=\|\bm{U}\|\|\bm{U}^{-1}\bm{b}\| $.
\end{theorem}
\begin{IEEEproof}
	Due to the error dynamics in \eqref{eq:err_dyn}, its solution is given by
	\begin{align}\label{eq:e_int1}
	\bm{e}(t) = \mathrm{e}^{\bm{A}_{\bm{\theta}}t}\bm{e}(0)+\int\limits_0^t \mathrm{e}^{\bm{A}_{\bm{\theta}} (t-t') }\bm{b} f_e(t')\mathrm{d}t',
	\end{align}
	where $f_e(t)=f(\bm{x}(t))-\mu(\bm{x}(t))$. Therefore, we directly obtain 
	\begin{align}
	\|\bm{e}(t)\|\leq \|\mathrm{e}^{\bm{A}_{\bm{\theta}}t}\bm{e}(0)\|+\int\limits_0^t \|\mathrm{e}^{\bm{A}_{\bm{\theta}} (t-t') }\bm{b}\| |\bar{f}_e(t')|\mathrm{d}t',
	\end{align}
	where $\bar{f}_e(t)$ can be any function such that $|f_e(t)|\leq \bar{f}_e(t)$.
	Using the eigendecomposition of $\bm{A}_{\bm{\theta}}=\bm{U}\bm{\Lambda}\bm{U}^{-1}$, it can be directly seen that 
	\begin{align}
	\|\mathrm{e}^{\bm{A}_{\bm{\theta}}t}\bm{b}\|\leq \|\bm{U}\|\|\bm{U}^{-1}\bm{b}\| \mathrm{e}^{\lambda_{\max}(\bm{A}_{\bm{\theta}})t}.
	\end{align}
	Hence, we obtain
	\begin{align}
	\|\bm{e}(t)\|\leq& \|\bm{U}\|\|\bm{U}^{-1}\bm{e}(0)\| \mathrm{e}^{\lambda_{\max}(\bm{A}_{\bm{\theta}})t}\nonumber\\
	&+\|\bm{U}\|\|\bm{U}^{-1}\bm{b}\|\int\limits_0^t \mathrm{e}^{\lambda_{\max}(\bm{A}_{\bm{\theta}}) (t-t') }  |f_e(t')|\mathrm{d}t'.
	\end{align}
	 The right handside of this inequality is again the solution of a differential equation such hat $\|\bm{e}(t)\|\leq \tilde{\upsilon}$ for 
	\begin{align}\label{eq:ups_tilde}
		\dot{\tilde{\upsilon}}=\lambda_{\max}(\bm{A}_{\bm{\theta}})\tilde{\upsilon}+\|\bm{U}\|\|\bm{U}^{-1}\bm{b}\| \bar{f}_e(t)
	\end{align}
	with $\tilde{\upsilon}(0)=\|\bm{U}\|\|\bm{U}^{-1}\bm{e}(0)\|$.
	It remains to derive a bound $\bar{f}_e(t)$ for $|f_e(t)|$ in \eqref{eq:ups_tilde}. Due to \cref{th:errbound_with}, it holds that $|f_e(t)|\leq \eta_N(\bm{x}(t))$ for all $\bm{x}\in\mathbb{X}$ with probability of at least $1-\delta$. Moreover, we have $\eta_N(\bm{x}(t))\leq \eta_N(\bm{x}_{\mathrm{ref}}(t))+L_{\sigma}\sqrt{\beta_{\mathbb{X}}(\tau)}\|\bm{e}(t)\|$ due to Lipschitz continuity of $\sigma(\cdot)$ guaranteed by \cref{cor:Lipschitz var stat}. Therefore, it follows that 
	\begin{align}
		\dot{\tilde{\upsilon}}\leq \left(\lambda_{\max}(\bm{A}_{\bm{\theta}})+L_{\sigma}\zeta\sqrt{\beta_{\mathbb{X}}(\tau)}\right)\tilde{\upsilon} + \zeta \eta(\bm{x}_{\mathrm{ref}}),
	\end{align}
	which concludes the proof.
\end{IEEEproof}

Since $\eta(\bm{x}_{\mathrm{ref}})$ can be directly computed at any time instant, determining the tracking error bound using \cref{th:tracking_time_varying} simply requires simulating the linear dynamical system \eqref{eq:xi_diff}. This can be straightforwardly done for a given time horizon in contrast to similar prior approaches \cite{Lederer2019, Umlauft2020}, where the uniform error bound needs to be determined at the actual system state $\bm{x}$. In order to achieve this improved practical applicability, additional requirements on the stability of the linear dynamics described by $\bm{A}_{\bm{\theta}}$ are necessary. It is obvious that \eqref{eq:ebound} only remains bounded if the linear dynamics \eqref{eq:xi_diff} are stable, which can be straightforwardly shown to require $\lambda_{\max}(\bm{A}_{\bm{\theta}})<-L_{\sigma}\zeta\sqrt{\beta_{\mathbb{X}}(\tau)}. $
Due to the dependency of the eigenvalue $\lambda_{\max}(\bm{A}_{\bm{\theta}})$ on the parameters $\bm{\theta}$, this condition can be satisfied if 
\begin{align}\label{eq:gain_cond}
	\|\bm{\theta}\|\geq \alpha^{-1}\left(-L_{\sigma}\zeta\sqrt{\beta_{\mathbb{X}}(\tau)}\right).
\end{align}
Therefore, this condition effectively poses a lower bound on the admissible control gains.

\subsection{Dependency of Accuracy Guarantees on Data Density}\label{subsec:data2tracke}

While \cref{th:tracking_time_varying} provides an accurate bound for the tracking error depending on the local data density, 
it is challenging to apply this result to the asymptotic analysis of the tracking error. Therefore, we bound the maximum tracking 
error along the reference trajectory as shown in the following proposition.
\begin{proposition}\label{prop:track_time_varying}
		Consider a linear system \eqref{eq:sys} satisfying \cref{as:stab}, which is perturbed by 
		a $L_f$-Lipschitz nonlinearity $f(\cdot)$ satisfying \cref{ass:samplefun}. Assume that a zero
		mean Gaussian process with $L_k$-Lipschitz stationary kernel is used to learn a model $\hat{f}(\cdot)=\mu(\cdot)$ of $f(\cdot)$,
		such that a controller~\eqref{eq:ctrl} is used to track the bounded 
		reference $\bm{x}_{\mathrm{ref}}$. If \eqref{eq:gain_cond} is satisfied, then, for $\|\bm{e}(0)\|=0$, the maximum tracking error is bounded by $\sup_{t\geq 0}\|\bm{e}(t)\|\leq \bar{\upsilon}$ with probability of at least $1-\delta$, where
		\begin{align}\label{eq:upsilon_bar}
			\bar{\upsilon} &= -\frac{\zeta }{\lambda_{\max}(\bm{A}_{\bm{\theta}})+L_{\sigma}\zeta\sqrt{\beta_{\mathbb{X}}(\tau)}}\sup\limits_{t\geq 0}\eta(\bm{x}_{\mathrm{ref}}(t)).
		\end{align}
\end{proposition}
\begin{IEEEproof}
	It immediately follows from \eqref{eq:e_int1} that 
	\begin{align}
	\|\bm{e}(t)\|&\leq\\ 
	&\zeta\int\limits_0^t \mathrm{e}^{\left(\lambda_{\max}(\bm{A}_{\bm{\theta}})+L_{\sigma}\zeta\sqrt{\beta_{\mathbb{X}}(\tau)}\right) (t-t') }  \mathrm{d}t'\sup\limits_{0\leq t'\leq t}\eta(\bm{x}_{\mathrm{ref}}(t')).\nonumber
	\end{align}
	Since the integral can be straightforwardly calculated, we obtain
	\begin{align}
		\sup\limits_{t\geq 0}\|\bm{e}(t)\|\leq -\frac{\zeta\sup\limits_{t\geq 0}\eta(\bm{x}_{\mathrm{ref}}(t))}{\lambda_{\max}(\bm{A}_{\bm{\theta}})+L_{\sigma}\zeta\sqrt{\beta_{\mathbb{X}}(\tau)}},
	\end{align}
	which concludes the proof.
\end{IEEEproof}
Note that the restriction to a zero initial condition is only considered to simplify the derivation, but the extension to non-zero initial conditions is straightforward. Therefore, the assumptions of \cref{prop:track_time_varying} are not more restrictive than those of \cref{th:tracking_time_varying}. 

In order to analyze the asymptotic behavior of the tracking error, we combine \cref{prop:track_time_varying} with \cref{prop:rho_decay}. Using the shorthand notation $\underline{\rho}=\inf_{t\geq 0} \rho(\bm{x}_{\mathrm{ref}}(t))$, this results in the following theorem.
\begin{theorem}\label{th:track_error_asymp}
	Consider a linear system \eqref{eq:sys} satisfying \cref{as:stab}, which is perturbed by 
	a $L_f$-Lipschitz nonlinearity $f(\cdot)$ satisfying \cref{ass:samplefun}. Assume that a zero
	mean Gaussian process with $L_k$-Lipschitz stationary kernel is used to learn a model $\hat{f}(\cdot)=\mu(\cdot)$ of $f(\cdot)$,
	such that a controller~\eqref{eq:ctrl} is used to track the bounded 
	reference $\bm{x}_{\mathrm{ref}}$.
	Choose $\tau$ such that $\beta_{\mathbb{X}}(\tau)\geq  \nicefrac{\gamma^2(\tau)\underline{\rho}k(\bm{0})}{2}$ and $\bm{\theta}$ such that 
	\begin{align}\label{eq:tilde_kappa}
		\kappa=-\frac{2\zeta\sqrt{\beta_{\mathbb{X}}(\tau)}}{\lambda_{\max}(\bm{A}_{\bm{\theta}})+L_{\sigma}\zeta\sqrt{\beta_{\mathbb{X}}(\tau)}}
	\end{align}
	is constant and \eqref{eq:gain_cond} is satisfied.
	Then, for $\|\bm{e}(0)\|=0$, the maximum tracking error bound asymptotically behaves as 
	\begin{align}\label{eq:rho2e}
	\bar{\upsilon} \in\mathcal{O}\left(\frac{1}{\sqrt{\underline{\rho}}}\right).
	\end{align}
\end{theorem}
\begin{IEEEproof}
%
	We first focus on the asymptotic behavior of the maximum learning error bound along the reference $\sup_{t\geq 0}\eta(\bm{x}_{\mathrm{ref}}(t))$, which can be expressed as
	\begin{align}
		\sup\limits_{t\geq 0}\eta(\bm{x}_{\mathrm{ref}}(t)) = \sqrt{\beta_{\mathbb{X}}(\tau)}\sup\limits_{t\geq 0}\sigma(\bm{x}_{\mathrm{ref}}(t))+\gamma(\tau).
	\end{align}
	Due to \cref{prop:rho_decay}, the considered parameter $\beta_{\mathbb{X}}(\tau)$ implies 
	\begin{align}\label{eq:tau_cond}
		\sup\limits_{t\geq 0}\sigma(\bm{x}_{\mathrm{ref}}(t))\geq \frac{\gamma(\tau)}{\sqrt{\beta_{\mathbb{X}}(\tau)}},
	\end{align}
	such that we can simplify the learning error bound to 
	\begin{align}\label{eq:unibound_without_gamma}
		\sup\limits_{t\geq 0}\eta(\bm{x}_{\mathrm{ref}}(t)) \leq 2\sqrt{\beta_{\mathbb{X}}(\tau)}\sup\limits_{t\geq 0}\sigma(\bm{x}_{\mathrm{ref}}(t)).
	\end{align}
	Therefore, it follows from  proposition \cref{prop:track_time_varying} that
	\begin{align}
		\bar{\upsilon} = \kappa\sup\limits_{t\geq 0}\sigma(\bm{x}_{\mathrm{ref}}(t)),
	\end{align}
	whose asymptotic behavior only depends on $\sigma(\bm{x}_{\mathrm{ref}}(t))$ due to the assumed constant value of $\tilde{\kappa}$, i.e., $\bar{\upsilon}\in\mathcal{O}(\sup_{t\geq 0}\sigma(\bm{x}_{\mathrm{ref}}(t)))$.	
	Due to  \cref{prop:rho_decay}, we have $\sup_{t\geq 0}\sigma(\bm{x}_{\mathrm{ref}}(t))\in\mathcal{O}(\nicefrac{1}{\sqrt{\underline{\rho}}})$,
	which concludes the proof.
\end{IEEEproof}
This theorem establishes a direct relationship between the minimum data density $\underline{\rho}$ along the reference trajectory $\bm{x}_{\mathrm{ref}}(t)$ and the maximum of the tracking error $\bm{e}$, showing that an arbitrarily small tracking error can be guaranteed when suitable data is available. Since this requires a vanishing $\gamma(\tau)$, $\beta_{\mathbb{X}}(\tau)$ must grow. The chosen $\beta_{\mathbb{X}}(\tau)$ in \cref{th:track_error_asymp} satisfies this property. In order to see this note that $\sqrt{\beta_{\mathbb{X}}(\tau)}$ is growing with decreasing $\tau$ and $\gamma(\tau)\in\mathcal{O}(N\tau)$ holds for stationary kernels. Therefore, we can set $\tau\propto \nicefrac{1}{(N\sqrt{\underline{\rho}})}$, which directly yields $\beta_{\mathbb{X}}(\tau)\propto \log(N\sqrt{\underline{\rho}})$. Due to condition \eqref{eq:tilde_kappa}, this increase rate of $\beta_{\mathbb{X}}(\tau)$ finally requires reducing eigenvalues $-\lambda_{\max}(\bm{A}_{\bm{\theta}})\propto \sqrt{\log(N\sqrt{\underline{\rho}})}$.
While this increase requirement might seem like a restrictive assumption, it is important to note that without learning, it follows from the proof of \cref{prop:track_time_varying} that $-\lambda_{\max}(\bm{A}_{\bm{\theta}})\propto \nicefrac{1}{\bar{\upsilon}}$. In contrast, we immediately obtain $\underline{\rho}\propto\nicefrac{1}{\bar{\upsilon}^2}$ from \eqref{eq:rho2e}, such that $-\lambda_{\max}(\bm{A}_{\bm{\theta}})\propto\sqrt{\log(\nicefrac{N}{\bar{\upsilon}})}$ holds. Assuming the number of training samples $N$ grows at most polynomially with $\underline{\rho}$ as ensured, e.g., for the case of SE or Mat\'ern kernels with uniformly distributed training data discussed in \cref{subsec:varbound_stat}, this finally implies $-\lambda_{\max}(\bm{A}_{\bm{\theta}})\in\mathcal{O}(\sqrt{\log(\nicefrac{1}{\bar{\upsilon}})})$. Therefore, the requirement on the growth rate for ensuring arbitrarily small tracking errors reduces from hyperbolic to log-hyperbolic with suitable training data.\looseness=-1

\subsection{Episodic Data Generation for Prescribed Performance}\label{subsec:epi_data}

Although \cref{th:track_error_asymp} provides conditions for training data to ensure an arbitrarily small tracking error $\bm{e}$, 
it does not provide direct insights how suitable training data sets can be obtained. Therefore, we develop an episodic approach for generating training data sets in this section. For simplicity, we consider a constant sampling time $T_s\in\mathbb{R}_+$ during each episode with execution time $T_p\in\mathbb{R}_+$, which yields data sets of the form
\begin{align}
\mathbb{D}_N^{T_s}=\left\{(\bm{x}(iT_s),f(\bm{x}(iT_s))+\epsilon^{(i)}) \right\}_{i=0}^{N_p},
\end{align}
where $N_p = \left\lfloor 1+\nicefrac{T_p}{T_s} \right\rfloor$ denotes the number of training samples gathered during one episode. Therefore, the tracking error bound $\bar{\upsilon}$ from one episode immediately provides guarantees for the training data of the next episode. We exploit this by adjusting the sampling time $T_s$ and the maximum eigenvalue $\lambda_{\max}(\bm{A}_{\bm{\theta}})$ as demonstrated in \cref{alg:itlearn} in order to ensure a sufficiently small error bound for the next episode. This dependency on the sampling time is emphasized by an index $T_s$ in the posterior standard deviation $\sigma_{T_s}(\cdot)$. As shown in the following theorem, this approach guarantees the termination of \cref{alg:itlearn} after a finite number of iterations.

\begin{algorithm}[t]
	\small 
	\SetInd{0.5em}{0.4em}
	\DontPrintSemicolon
	\SetKwFunction{FSOP}{LearnControl}
	\SetKwFunction{FRes}{Resample}
	\SetKwProg{Fn}{Function}{:}{}
	\Fn{\FSOP{$\bar{e}$}}{
	$\mu(\cdot)\gets 0$, $\sigma^2(\cdot)\gets \sigma_f^2$, $i\gets0$\;
	compute $\bar{\upsilon}_0$ using \eqref{eq:upsilon_bar} for $\tau$ satisfying \eqref{eq:tau_cond}\;
	\While{$\bar{\upsilon}_i>\bar{e}$}{
		Initialize system at $\bm{x}(0)=\bm{x}_{\mathrm{ref}}(0)$, $i\gets i+1$\;
		\While{$t\leq T_p$}{
			run controller \eqref{eq:ctrl} on system \eqref{eq:sys}\;
		}
		choose $\bm{\theta}$ such that \eqref{eq:ktilde_cond} holds\;
		find $T_s$ such that \eqref{eq:Ts_cond} is satisfied\;
		determine $\mu_{T_s}(\cdot)$ and $\sigma^2_{T_s}(\cdot)$ with $\mathbb{D}_N^{T_s}$ using \eqref{eq:mean}, \eqref{eq:var}\;
		compute $\bar{\upsilon}_i$ using \eqref{eq:upsilon_bar} for $\tau$ satisfying \eqref{eq:tau_cond}\;
	}
		
	}
%
	\caption{Iterative Learning for Asymptotic Stability}
	\label{alg:itlearn}
\end{algorithm}


%

\begin{theorem}\label{th:convergence}	
	Consider a linear system \eqref{eq:sys} satisfying \cref{as:stab}, which is perturbed by 
	a $L_f$-Lipschitz nonlinearity $f(\cdot)$ satisfying \cref{ass:samplefun}. Assume that a zero
	mean Gaussian process with $L_k$-Lipschitz stationary kernel is used to learn a model $\hat{f}(\cdot)=\mu(\cdot)$ of $f(\cdot)$,
	such that a controller~\eqref{eq:ctrl} is used to track the bounded 
	reference $\bm{x}_{\mathrm{ref}}$. If $\bm{\theta}$ and $T_s$ are chosen such that
	\begin{align}\label{eq:ktilde_cond}
		-\lambda_{\max}(\bm{A}_{\bm{\theta}})&\geq \frac{8\sqrt{L_{\partial k}} )+\xi L_{\sigma}}{\xi}\zeta\sqrt{\beta_{\mathbb{X}}(\tau)}\\
		\max_{0\leq t \leq T_p}\sigma^2_{T_s}(\bm{x}_{\mathrm{ref}}(t))&\leq 16L_{\partial k}\bar{\upsilon}_{i-1}^2
		\label{eq:Ts_cond}
	\end{align}
	holds in every episode for $\xi<1$, 
	\cref{alg:itlearn} terminates after at most
	\begin{align}
		N_E=\left\lceil\frac{\log\left(4\bar{e}\sqrt{L_{\partial k}}\right)-\log\left(\sqrt{k(\bm{0})}\right)}{\log(\xi)}\right\rceil
	\end{align}
	episodes with probability of at least $1-N_E\delta$. 
\end{theorem}
\begin{IEEEproof}
	It is straightforward to see that \eqref{eq:Ts_cond} together with \cref{prop:track_time_varying} implies
	\begin{align}
		\bar{\upsilon}_0&=\kappa \sqrt{k(\bm{0})},\\
		\bar{\upsilon}_{i+1}&=4\sqrt{L_{\partial k}}\kappa \bar{\upsilon}_i
	\end{align}
	for $\tau$ such that \eqref{eq:tau_cond} is satisfied, 
	where the index $i$ is used to denote the episode. 
	Since $4\sqrt{L_{\partial k}}\kappa\leq\xi<1$ holds due to \eqref{eq:ktilde_cond}, it immediately follows that $\bar{\upsilon}_i$ decays exponentially, i.e., $\bar{\upsilon}_i=\xi^i\bar{\upsilon}_0$ with probability of at least $1-\delta$ for each episode. Therefore, \cref{alg:itlearn} is guaranteed to terminate after $N_E$ episodes with probability of at least $1-N_E\delta$ due to the union bound. \looseness=-1
\end{IEEEproof}
Due to the exponential decay of the tracking error bound $\bar{\upsilon}$ ensured by \cref{th:convergence}, \cref{alg:itlearn} quickly terminates. This comes at the price of higher requirements \eqref{eq:ktilde_cond} on the eigenvalues of $\bm{A}_{\bm{\theta}}$ compared to \cref{prop:track_time_varying}. However, the difference is merely a constant factor, and it is indeed straightforward to see that $-\lambda_{\max}(\bm{A}_{\bm{\theta}})\propto \nicefrac{1}{\sqrt{\log(\bar{e})}}$ is sufficient to compensate the effect of an increasing $\beta_{\mathbb{X}}(\tau)$ for all polynomially growing data sets. Therefore, this requirement is still significantly lower compared to ensuring the tracking error bound $\bar{e}$ without learning as discussed in \cref{subsec:data2tracke}. 

While the results in previous sections posed requirements on the data distribution in terms of the data density $\rho(\bm{x})$, \cref{th:convergence} explicitly considers the data generation process by providing an upper bound for the sampling time $T_s$ in \eqref{eq:Ts_cond}. Due to the form of this condition, it cannot be computed before the controller is applied to the system, but it can easily be verified a posteriori. Therefore, we can ensure it via a sufficiently high sampling rate during the application of the controller, such that we simply can downsample the obtained data to the necessary sampling time $T_s$. The required maximum sampling rate can be bounded using the following proposition.
\begin{proposition}
	Consider a linear system \eqref{eq:sys} satisfying \cref{as:stab}, which is perturbed by 
	a $L_f$-Lipschitz nonlinearity $f(\cdot)$ satisfying \cref{ass:samplefun}. Assume that a zero
	mean Gaussian process with $L_k$-Lipschitz stationary kernel is used to learn a model $\hat{f}(\cdot)=\mu(\cdot)$ of $f(\cdot)$,
	such that a controller~\eqref{eq:ctrl} is used to track the continuous, bounded 
	reference $\bm{x}_{\mathrm{ref}}$. 
		Then, the sampling time $T_s$ required by condition \eqref{eq:Ts_cond} in \cref{alg:itlearn} is bounded by
		\begin{align}
			T_s\geq \underline{T}_s=\frac{16L_{\partial k}\bar{e}^3}{\sigma_{\mathrm{on}}^2\max_{0\leq t \leq T_p}\! \|\dot{\bm{x}}(t)\|}.
		\end{align}
\end{proposition}
\begin{IEEEproof}
	We prove this proposition by deriving a value of $T_s$ which satisfies \eqref{eq:Ts_cond}
	Due to \cref{prop:rho_decay}, \eqref{eq:Ts_cond} is guaranteed to hold if $\underline{\rho}\geq \nicefrac{1}{(8L_{\partial k}\sigma_f^2 \bar{\upsilon}_{i-1}^2)}$. Set $\rho'=\nicefrac{1}{(8L_{\partial k}\sigma_f^2 \bar{\upsilon}_{i-1}^2)}$. Then, it follows from \cref{lem:K_to_Euclid} that
	\begin{align}
	\mathbb{B}_{2\upsilon_{i-1}}(\bm{x}_{\mathrm{ref}}(t))\subset\mathbb{K}_{\rho'}(\bm{x}_{\mathrm{ref}}(t)).
	\end{align}
	The Euclidean ball around $\bm{x}_{\mathrm{ref}}(t)$ on the left handside can be inner bounded by a Euclidean ball with half the radius around the actual trajectory, i.e., 
	\begin{align}
	\mathbb{B}_{\bar{\upsilon}_{i-1}}(\bm{x}(t))\subset \mathbb{B}_{2\bar{\upsilon}_{i-1}}(\bm{x}_{\mathrm{ref}}(t)).
	\end{align}
	The smaller Euclidean ball has a diameter of $\bar{\upsilon}_{i-1}$ and the actual trajectory passes through its center. Moreover, the distance between two samples can be bounded by $T_s \max_{0\leq t \leq T_p} \|\dot{\bm{x}}(t)\|$. Note that the maximum temporal derivative of the state is bounded. In order to see this, note that we can express the dynamics of the system as
	\begin{align}
		\dot{\bm{x}}=\dot{\bm{x}}_{\mathrm{ref}}+\bm{A}_{\bm{\theta}}\bm{e}+\bm{b}(f(\bm{x})-\mu(\bm{x}).
	\end{align}
	Due to the bounded prediction error, the bounded tracking error and the continuous reference trajectory, we can therefore bound the state derivative by
	\begin{align}
		\max\limits_{0\leq t\leq T_p} \!\|\dot{\bm{x}}(t)\|&\leq \left(\!\|\bm{A}_{\bm{\theta}}\|\!+\!\sqrt{\beta_{\mathbb{X}}(\tau)}L_{\sigma}\!\right)\!\bar{\upsilon}_i\!+\!\max\limits_{0\leq t\leq T_p}\!\eta(\bm{x}_{\mathrm{ref}}(t))\nonumber\\
		&+\max\limits_{0\leq t\leq T_p}\! \|\dot{\bm{x}}_{\mathrm{ref}}(t)\|.
	\end{align}
	This allows us to bound the number of points in $\mathbb{K}_{\rho'}(\bm{x}_{\mathrm{ref}}(t))$ by
	\begin{align}
	\!\!|\mathbb{K}_{\rho'}(\bm{x}_{\mathrm{ref}}(t))|\!\geq\! |\mathbb{B}_{\bar{\upsilon}_{i-1}}(\bm{x}(t))|\!\geq\! \frac{2\bar{\upsilon}_{i-1}}{
		T_s\max_{0\leq t \leq T_p}\! \|\dot{\bm{x}}(t)\|}\!.\!
	\end{align}
	For $\underline{\rho}\geq \rho'$, it must hold that
	\begin{align}
	\frac{2\upsilon_{i-1}}{
		T_s\max_{0\leq t \leq T_p} \|\dot{\bm{x}}(t)\|}\geq \rho'\sigma_{\mathrm{on}}^2k(\bm{0})=\frac{\sigma_{\mathrm{on}}^2}{8L_{\partial k} \upsilon_{i-1}^2}
	\end{align}
	due to \eqref{eq:kernel_neighborhood}. This inequality can be ensured to hold by setting 
	\begin{align}
	T_s=\frac{16L_{\partial k}\bar{\upsilon}^3_{i-1}}{\sigma_{\mathrm{on}}^2\max_{0\leq t \leq T_p}\! \|\dot{\bm{x}}(t)\|},
	\end{align}
	which concludes the proof.
\end{IEEEproof}


\section{Numerical Evaluation}
\label{sec:numEval}

In order to demonstrate the flexibility and effectiveness of the derived theoretical results, we compare the tracking error bounds with empirically observed tracking errors in different simulations. In \cref{subsec:safety}, we evaluate the time-varying 
tracking error bound for training data unevenly distributed over the relevant part of the state space $\mathbb{X}$. The behavior of the asymptotic error bound is investigated in \cref{subsec:sim_data2e}. Finally, we demonstrate the effectiveness of the proposed episodic data generation approach for ensuring a desired tracking accuracy in \cref{subsec:sim_episodic}.

\subsection{Data-dependency of Safety Regions}
\label{subsec:safety}

For evaluating the time-varying tracking error bound, we consider a 
nonlinear dynamical system 
\begin{align}
\dot{x}_1=x_2,&&\dot{x}_2=f(\bm{x})+g(\bm{x})u,
\end{align}
where \mbox{$f(\bm{x}) = 1-\sin(2x_1) +  \nicefrac{1}{(1+\exp(-x_2)})$} and $g(\bm{x})= 1+\nicefrac{1}{2}\sin\left(\nicefrac{x_2}{2}\right)$, which is a marginal variation of the system considered in \cite{Umlauft2020}.
Assuming exact knowledge of $\bm{g}(\cdot)$, we can approximately feedback linearize this system and apply a linear tracking controller $u_{\mathrm{lin}}=-\theta_1\theta_2 x_2-\theta_2 x_2$, where $\theta_1,\theta_2\in\mathbb{R}_+$ are 
design parameters. This yields a 
two-dimensional system of the form~\eqref{eq:sys} 
with
\begin{align}
	\bm{A}_{\bm{\theta}}=\begin{bmatrix}
	0&1\\ -\theta_1\theta_2& -\theta_2
	\end{bmatrix}&&\bm{b}=\begin{bmatrix}
	0\\1
	\end{bmatrix}.
\end{align}
In order to demonstrate the effect of the distribution, we use a uniform 
grid over~$[0\ 3]\times[-4\ 4]$ with~$25$ points and~$\sigma_{\mathrm{on}}^2 = 0.01$ as 
training data set, such that half of the considered state 
space~$\mathbb{X} =[-5\ 5]^2$ is not covered by training data.
A SE kernel with automatic 
relevance determination is employed for Gaussian process regression
and the hyperparameters are optimized using likelihood maximization.
For computing the uniform prediction error bound in 
\cref{th:errbound_with}, we 
set $\tau=0.01$, $\delta=0.01$ and $L_f=2$. 
The task is to track the circular reference trajectory~$x_d(t) = 2\sin(t)$ with state
$x_1$, which leads to the reference trajectory $\bm{x}_{\mathrm{ref}}(t)=[2\sin(t)\ 2\cos(t)]^T$. 
We aim to achieve this using~$\theta_1=10$ and~$\theta_2=20$, which can be shown to satisfy condition
\eqref{eq:gain_cond}.

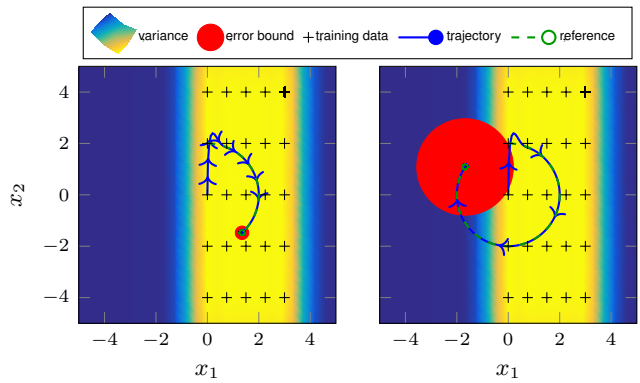
\begin{figure}[t]

		\centering 
		\def\file{files/2D.txt}	 \def\imax{177} \def\ifin{1000}
		\def\imin{81}
		\def\filesig{files/2D_var.txt}
		\tikzsetnextfilename{LyapDecrTe_max}
		\pgfplotsset{
			colormap={parula}{
				rgb255=(249,251,14)
				rgb255=(252,206,46)
				rgb255=(225,185,82)
				rgb255=(165,190,107)
				rgb255=(89,189,140)
				rgb255=(21,177,180)
				rgb255=(7,156,207)
				rgb255=(18,125,216)
				rgb255=(15,92,221)
				rgb255=(53,42,135)	
			},
		}
		\begin{tikzpicture}
		\begin{axis}[mesh/rows=25,
		grid=none,enlargelimits=false, axis on top,
		xlabel={$x_1$}, ylabel={$x_2$}, legend pos = north west,
		xmin=-5, xmax = 5, ymin = -5, ymax = 5,	legend columns=5,
		height=5cm, width=5cm, 
		]
		\addplot[surf,shader=interp, samples= 2, point meta rel=per plot, point meta=explicit] table[x=Xtes_1,y=Xtes_2,meta=sig2]{\filesig};
		\addplot[red,fill=red,forget plot] table[x=Xhullmin_1,y=Xhullmin_2]{\file};
		\addplot[only marks,black, mark=+]	table[x = Xtr_1, y = Xtr_2] {\file};
		\addplot[blue,only marks, mark size=1pt, forget plot] table[x = X_min_1, y = X_min_2]{\file};
		\addplot[forget plot,thick,blue,postaction={decorate,decoration={markings,
				mark=between positions 0.1 and 0.9 step 0.1 with {\arrow{>}}}}]	
		table[x=Xsim_1,y=Xsim_2,skip coords between 
		index={\imin}{\ifin}]{\file};
		\addplot[green!60!black,only marks, mark size=1pt, forget plot, mark=o] table[x = Xd_min_1, y = Xd_min_2]{\file};
		\addplot[dashed,green!60!black,thick]	
		table[x=Xd_1,y=Xd_2, skip coords between index={\imin}{\ifin}] {\file};
		\end{axis}

		\begin{axis}[mesh/rows=25,grid=none,enlargelimits=false, axis on top, 
		legend columns=6,
		xlabel={$x_1$}, yticklabels={,,},
		legend style={at={(-1.155,1.13)},anchor=west},
		xmin=-5, xmax = 5, ymin = -5, ymax = 5,
		height=5cm, width=5cm, xshift=4cm,
		]
		\addplot[surf,shader=interp, samples= 2, point meta rel=per plot, point meta=explicit] table[x=Xtes_1,y=Xtes_2,meta=sig2]{\filesig};
		\addlegendentry{$\!\!$variance$~$};
		\addplot[red,fill=red, forget plot] table[x=Xhullmax_1,y=Xhullmax_2]{\file};
		\addlegendimage{only marks, mark=*,color=red,mark options={scale=2.5}}
		\addlegendentry{$\!\!$error bound$~$};
		\addplot[only marks,black, mark=+]	table[x = Xtr_1, y = Xtr_2] {\file};
		\addlegendentry{$\!\!$training data$~$};
		\addplot[blue,only marks, mark size=1pt, forget plot] table[x = X_max_1, y = X_max_2]{\file};
		\addplot[blue,thick,forget plot, postaction={decorate,decoration={markings,
				mark=between positions 0.1 and 0.9 step 0.2 with {\arrow{>}}}}]	
		table[x=Xsim_1,y=Xsim_2,skip coords between 
		index={\imax}{\ifin}]{\file};
		\addlegendimage{blue,-*, thick};
		\addlegendentry{$\!\!$trajectory$~$};
		\addplot[green!60!black,only marks, mark size=1pt, forget plot, mark=o] table[x = Xd_max_1, y = Xd_max_2]{\file};
		\addplot[dashed,green!60!black, thick, forget plot]	
		table[x=Xd_1,y=Xd_2,skip coords between index={\imax}{\ifin}] {\file};
		\addlegendimage{dashed,green!60!black,-o, thick};
		\addlegendentry{$\!\!$reference$~$};
		
		\end{axis}
		\end{tikzpicture}
	\caption{Reference trajectory and simulated trajectory together with the 
		illustration of the tracking error bound $\upsilon(t)$. Low posterior standard 
		deviations lead to significantly smaller tracking error bounds.}
	\label{fig:LyapDecrTe}
\end{figure}

Snapshots of the resulting trajectory together with visualizations of the tracking error bounds
obtained using \cref{th:tracking_time_varying} are illustrated in \cref{fig:LyapDecrTe}. When
the GP standard deviation $\sigma(\bm{x}_{\mathrm{ref}})$ is large, the tracking error 
bound $\upsilon(t)$ starts to increase, such that it reaches its maximum just before the system 
enters the region with low standard deviation. Afterwards, the feedback 
controller reduces the tracking error until the standard deviation starts
to increase again. This leads to the minimum tracking error bound illustrated on the left of 
\cref{fig:LyapDecrTe}. 

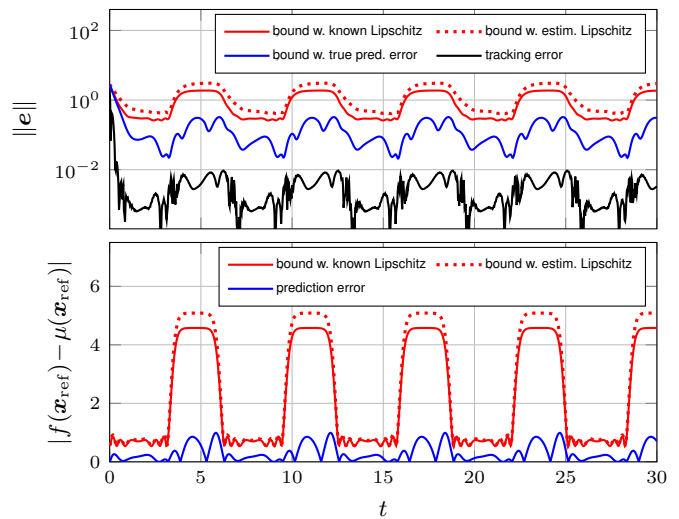
\begin{figure}[t]
	\center 
	\def\file{files/2D.txt}	\tikzsetnextfilename{error}
	\begin{tikzpicture}
	\begin{semilogyaxis}[ylabel={$\|\bm{e}\|$},xticklabels={,,},
	xmin=0, ymin = 0.0002, xmax = 30,ymax=400,legend columns=2,
	width=\columnwidth,height=4.5cm,
	]
	\addplot[red, thick]	table[x = T, y  = eb_GPbound]{\file};
	\addplot[red, dotted, very thick]	table[x = T, y  = eb_GPbound_Lfh]{\file};
	\addplot[blue, thick]	table[x = T, y  = eb_error]{\file};
	\addplot[black, thick]	table[x = T, y  = norme]{\file};
	
	\legend{$\!\!$bound w. known Lipschitz$~~$, $\!\!$bound w. estim. Lipschitz, $\!\!$bound w. true pred. error$~~$,  $\!\!$tracking error}
	\end{semilogyaxis}
	\begin{axis}[xlabel={$t$},ylabel={$|f(\bm{x}_{\mathrm{ref}})\!-\!\mu(\bm{x}_{\mathrm{ref}})|$},
	xmin=0, ymin = 0, xmax = 30,ymax=7.5,legend columns=2,
	width=\columnwidth,height=4.5cm, yshift =-3.1cm, 
	]
	\addplot[red, thick]	table[x = T, y  = efb]{\file};
	\addplot[red, dotted, very thick]	table[x = T, y  = efb_Lfh]{\file};
	\addplot[blue, thick]	table[x = T, y  = ef]{\file};
	
	\legend{$\!\!$bound w. known Lipschitz$~$, $\!\!$bound w. estim. Lipschitz, $\!\!$prediction error}
	\end{axis}
	\end{tikzpicture}
	\caption{Top: Tracking error bounds computed using \eqref{eq:xi_diff} with different prediction error bounds as inputs in comparison to the observed tracking error. Bottom: Prediction error bounds in comparison to the true model error.}
	\label{fig:err_time_var}
\end{figure}

This effect can also be seen at the observed tracking error as illustrated in \cref{fig:err_time_var}, which has its peaks at times when the tracking error bound $\upsilon$ is large. Therefore, the tracking error bound $\upsilon$ reflects the behavior of the observed error $\|\bm{e}\|$ well, even though it is rather conservative. The sources of this conservatism can be easily investigated by determining the bound obtained when using the true model error $|f(\bm{x}_{\mathrm{ref}})\!-\!\mu(\bm{x}_{\mathrm{ref}})|$
as input in \eqref{eq:xi_diff}.
It is clearly visible that even with the knowledge of the true prediction error, the tracking error bound 
exhibits some conservatism due the linearization around the reference trajectory $\bm{x}_{\mathrm{ref}}$.
The remaining conservatism is a consequence of the prediction error bound $\eta(\bm{x}_{\mathrm{ref}})$ as visualized at the bottom of \cref{fig:err_time_var}. 
Even though this bound reflects the availability of data well, it needs to capture the probabilistic worst case and is therefore considerably larger than the actual prediction error $|f(\bm{x}_{\mathrm{ref}})\!-\!\mu(\bm{x}_{\mathrm{ref}})|$. This leads to the fact that the tracking error bound $\upsilon$ conservatively reflects the behavior of the observed tracking error $\|\bm{e}\|$. 
Note that the usage of a probabilistic Lipschitz constant $\hat{L}_f$ obtained via \cref{th:errbound_without} 
does not significantly change this behavior. The corresponding tracking error bound merely
becomes slightly larger since we can compensate the conservative value of $\hat{L}_f$ using a smaller value $\tau=10^{-3}$. Therefore, \cref{th:errbound_without} enables the effective computation
of prediction error bounds without knowledge of a Lipschitz constant of the unknown function $f(\cdot)$.

\subsection{Dependency of the Tracking Accuracy on the Data Density}\label{subsec:sim_data2e}

In order to investigate the dependency of the tracking error bound $\upsilon$ on the data density $\underline{\rho}$ in more detail, we consider the same setting as in \cref{subsec:safety}, but use grids with different grid constants defined on $[-4,4]^2$ as training data sets, such that they cover the whole relevant domain. Due to the varying size of the training data set, we determine $\tau$ by finding the maximum value satisfying \eqref{eq:tau_cond} using a line search. We set $\theta_1=\theta_2=\theta$, such that we can compute a gain $\theta$ ensuring $\kappa=10$ in \eqref{eq:tilde_kappa} for the obtained value of $\tau$.

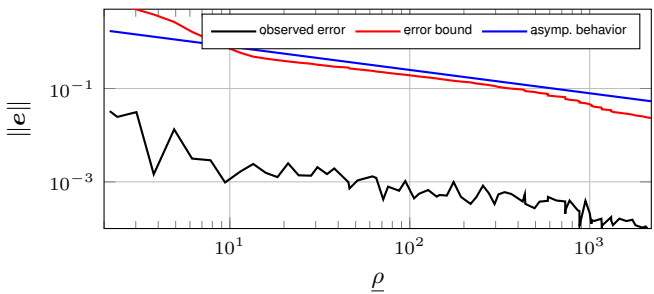
\begin{figure}[t]
	\center
	\centering 
	\centering 
	\def\file{files/error_asymptotics.txt}	\tikzsetnextfilename{error_asym}
	\begin{tikzpicture}
		\begin{loglogaxis}[xlabel={$\underline{\rho}$},ylabel={$\|\bm{e}\|$},
		xmin=2, ymin = 0.0001, xmax = 2200,ymax=5,legend columns=3,
		width=\columnwidth,height=4.5cm]
		\addplot[black, thick]	table[x = rho, y  = e]{\file};
		\addplot[red, thick]	table[x = rho, y  = eb]{\file};
		\addplot[blue, thick]	table[x = rho, y  = eb_asymp]{\file};
		\legend{$\!\!$observed error$~$, $\!\!$error bound$~$, $\!\!$asymp. behavior}
		\end{loglogaxis}
	\end{tikzpicture}
	
	\caption{Comparison of the observed tracking error, the tracking error bound and its guaranteed asymptotic decay rate for growing data densities $\underline{\rho}$.}
	\label{fig:error_asymp}
\end{figure}

\begin{figure}[t]
	\center
	\centering 
	\centering 
	\def\file{files/gain_asymptotics.txt}	\tikzsetnextfilename{gain_asym}
	\begin{tikzpicture}
	\begin{axis}[xlabel={$\nicefrac{1}{\sup_{t\geq0}\upsilon(t)}$},ylabel={$-\lambda_{\max}(\bm{A}_{\bm{\theta}})$},
	xmin=0, ymin = 0, xmax = 34.3,ymax=30,legend columns=3,
	width=\columnwidth,height=4.5cm]
	\addplot[red,thick]	table[x = eb_inv, y  = lambda]{\file};
	\addplot[blue, thick]	table[x = eb_inv, y  = lambda_asymp]{\file};
	\addplot[green]	table[x = eb_inv, y  = lambda_gain]{\file};
	\legend{$\!\!$GP controller$~$, $\!\!$asymp. behavior$~$,$\!\!$without compensation}
	\end{axis}
	\end{tikzpicture}
	
	\caption{Maximum eigenvalue $\lambda_{\max}(\bm{A}_{\bm{\theta}})$ necessary to ensure a given tracking error bound $\sup_{t\geq0}\upsilon(t)$ when learning a control law using equidistant grids in comparison to a pure feedback controller without compensation of nonlinearities.}
	\label{fig:kc_asymp}
\end{figure}
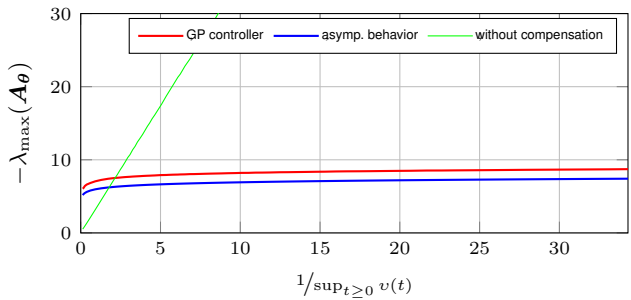

The resulting tracking errors $\|\bm{e}\|$ and bounds $\sup_{t\geq0}\upsilon(t)$ obtained with \cref{th:tracking_time_varying} for different data densities $\underline{\rho}$ are illustrated in \cref{fig:error_asymp}. Moreover, the asymptotic decay rate of $\bar{\upsilon}$ guaranteed by \cref{th:track_error_asymp} is depicted. It can be clearly seen that the asymptotic decay rate closely reflects the actual decay rate of the error bound $\sup_{t\geq0}\upsilon(t)$. Analogously to \cref{subsec:safety}, the tracking error bound is rather conservative, but the 
observed error $\|\bm{e}\|$ exhibits a decay rate with high similarity to its bound $\sup_{t\geq0}\upsilon(t)$. Despite this conservatism, the necessary  maximum eigenvalues $\lambda_{\max}(\bm{A}_{\bm{\theta}})$ for ensuring a low desired tracking error bound $\sup_{t\geq0}\upsilon(t)$ with such training data are significantly 
larger than without a controller compensating the nonlinearity as depicted in \cref{fig:kc_asymp}. This baseline comparison can be straightforwardly obtained as $\lambda_{\max}(\bm{A}_{\bm{\theta}})\geq \nicefrac{\zeta\bar{f}}{\bar{e}}$
by slightly adapting the proof of \cref{prop:track_time_varying} using $|f(\bm{x})|\leq\bar{f}$ and $\mu(\bm{x})=0$. Due to the linear growth of this 
condition with $\nicefrac{1}{\bar{e}}$, it quickly exceeds the maximum eigenvalue $\lambda_{\max}(\bm{A}_{\bm{\theta}})$ ensuring the same tracking error bound through the learned controller, even though we use 
the non-conservative bound $\bar{f}=3$. This clearly demonstrates the benefits of the derived theoretical results.

\subsection{Episodic Data Generation}\label{subsec:sim_episodic}

For evaluating the episodic data generation using \cref{alg:itlearn}, we consider the same setting as in \cref{subsec:sim_episodic}. Moreover, we set $\theta_1=\theta_2=\theta$ analogously to the previous section and choose $\theta$ such that $\xi=0.95$ holds in every iteration. 
A high frequency data set with sampling time $3\cdot 10^{-4}$ is generated in every episode, such that a line search can be used to determine the maximum value of $T_s$ satisfying \eqref{eq:Ts_cond}.

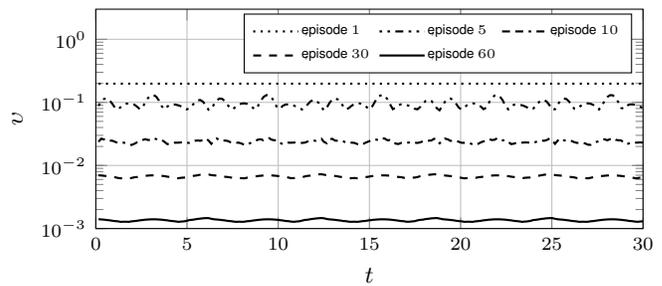
\begin{figure}[t]
	\center
	\centering 
	\centering 
	\def\file{files/error_examples_episodic.txt}	\tikzsetnextfilename{error_examples_episodic}
	\begin{tikzpicture}

	\begin{semilogyaxis}[xlabel={$t$},ylabel={$\upsilon$},
	xmin=0, ymin = 0.001, xmax = 30,ymax=3,legend columns=3,
	width=\columnwidth,height=4.5cm
	]
	\addplot[ thick, dotted]	table[x = t, y  = eb1]{\file};
	\addplot[ thick, dash pattern={on 1pt off 2pt on 1 pt off 2pt on 3pt off 2pt}]	table[x = t, y  = eb5]{\file};
	
	\addplot[ thick, dash pattern={on 3pt off 2pt on 3pt off 2pt on 1pt off 2pt}]	table[x = t, y  = eb10]{\file};
	\addplot[ thick, dashed]	table[x = t, y  = eb30]{\file};
	
	\addplot[thick]	table[x = t, y  = eb60]{\file};
	\legend{$\!\!$episode $1$$~$, $\!\!$episode $5$$~$, $\!\!$episode $10$,$\!\!$episode $30$$~$, $\!\!$episode $60$}
	\end{semilogyaxis}
	\end{tikzpicture}
	
	\caption{Tracking error bounds $\upsilon(t)$ for different episodes of \cref{alg:itlearn}.}
	\label{fig:error_examples}
\end{figure}

\begin{figure}[t]
	\center
	\centering 
	\centering 
	\def\file{files/error_episodic.txt}	\tikzsetnextfilename{error_episodic}
	\begin{tikzpicture}
	\begin{semilogyaxis}[xlabel={$N_E$},ylabel={$\|\bm{e}\|$},
	xmin=0, ymin = 0.00001, xmax = 69,ymax=1,legend columns=3,
	width=\columnwidth,height=4.5cm]
	\addplot[black, thick]	table[x = iteration, y  = e]{\file};
	\addplot[red, thick]	table[x = iteration, y  = eb]{\file};
	\addplot[blue, thick]	table[x = iteration, y  = eb_asymp]{\file};
	\legend{$\!\!$observed error$~$, $\!\!$error bound$~$, $\!\!$guaranteed decrease}
	\end{semilogyaxis}
	\end{tikzpicture}
	
	\caption{Decay rate of the tracking error bound $\sup_{t\geq0}\upsilon(t)$ and the observed tracking error $\|\bm{e}\|$ resulting from 
		\cref{alg:itlearn}.}
	\label{fig:error_epi}
\end{figure}
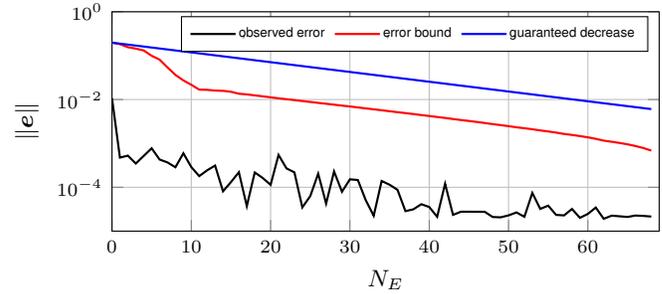

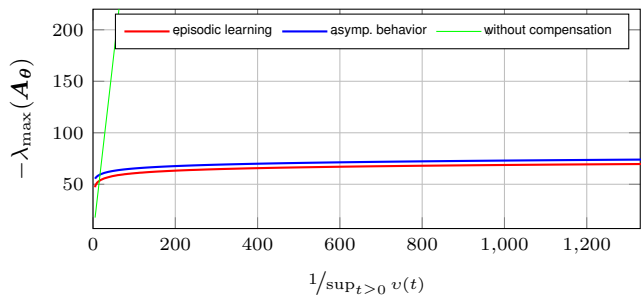
\begin{figure}[t]
	\center
	\centering 
	\centering 
	\def\file{files/gain_episodic.txt}	\tikzsetnextfilename{gain_episodic}
	\begin{tikzpicture}
	\begin{axis}[xlabel={$\nicefrac{1}{\sup_{t\geq0}\upsilon(t)}$},ylabel={$-\lambda_{\max}(\bm{A}_{\bm{\theta}})$},
	xmin=0, ymin = 7, xmax = 1330,ymax=220,legend columns=3,
	width=\columnwidth,height=4.5cm]
	\addplot[red,thick]	table[x = eb_inv, y  = lambda]{\file};
	\addplot[blue, thick]	table[x = eb_inv, y  = lambda_asymp]{\file};
	\addplot[green]	table[x = eb_inv, y  = lambda_gain]{\file};
	\legend{$\!\!$episodic learning$~$, $\!\!$asymp. behavior$~$, $\!\!$without compensation$~$};

	\end{axis}
	\end{tikzpicture}
	
	\caption{Maximum eigenvalue $\lambda_{\max}(\bm{A}_{\bm{\theta}})$ necessary to ensure a given tracking error bound $\sup_{t\geq0}\upsilon(t)$ when learning a control law using \cref{alg:itlearn} in comparison to a pure feedback controller without compensation of nonlinearities.}
	\label{fig:kc_epi}
\end{figure}

\begin{figure}[t]
	\center
	\centering 
	\centering 
	\def\file{files/Ts_episodic.txt}	\tikzsetnextfilename{Ts_episodic}
	\begin{tikzpicture}
	\begin{semilogyaxis}[xlabel={$\nicefrac{1}{\sup_{t\geq0}\upsilon(t)}$},ylabel={$T_s$},
	xmin=0, ymin = 1e-11, xmax = 1330,ymax=5,legend columns=3,
	width=\columnwidth,height=4.5cm]
	\addplot[red,thick]	table[x = eb_inv, y  = Ts]{\file};
	\addplot[blue, thick]	table[x = eb_inv, y  = Ts_asymp]{\file};
	\legend{$\!\!$used sampling time$~$, $\!\!$lower bound$~$}
	\end{semilogyaxis}
	\end{tikzpicture}
	
	\caption{Employed sampling time $T_s$ together with its lower bound $\underline{T}_s$ for given error bounds $\sup_{t\geq0}\upsilon(t)$ when running \cref{alg:itlearn}.}
	\label{fig:Ts_epi}
\end{figure}
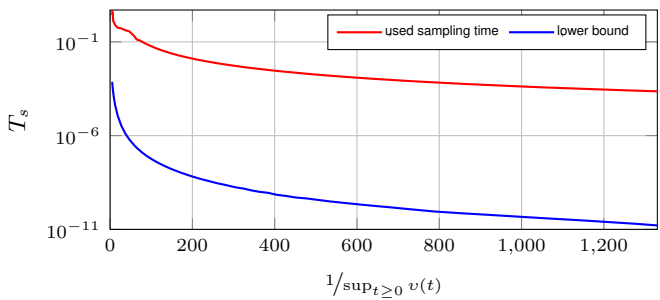

The tracking error bounds obtained form \cref{alg:itlearn} with these parameters are exemplarily illustrated for several
different episodes in \cref{fig:error_examples}. Due to the constant sampling time, the training data density along the 
reference is very similar within an episode, which directly leads to the rather minor variations in the tracking error bound
over time. Moreover, it can be seen that decrease of the tracking error bound $\upsilon$ is significantly larger during the first few episodes, before it slows down. This becomes even clearer when plotting the behavior of the error bound over the number of episodes as depicted in \cref{fig:error_epi}. During the first $10$ episodes the error bound $\sup_{t\geq0}\upsilon(t)$ decays faster than the guaranteed  rate of $\xi^{N_E}\bar{\upsilon}_0$, which is guaranteed by \cref{th:convergence}. This can be attributed to the fact that even a single additional data point reduces the posterior variance more than required for \eqref{eq:Ts_cond} at the beginning. Once a sufficiently large number of additional training samples is necessary to ensure \eqref{eq:Ts_cond}, this inaccuracy is overcome and the error bound $\sup_{t\geq0}\upsilon(t)$ closely follows the guaranteed decrease rate. In fact, the tracking error bound $\sup_{t\geq0}\upsilon(t)$, while being rather conservative similar to the
previous simulations, even reflects the behavior of the actually observed tracking error $\|\bm{e}\|$ accurately after $10$ episodes.

Note that this unexpected fast decay at the beginning has no influence on the 
required maximum eigenvalues $\lambda_{\max}(\bm{A}_{\bm{\theta}})$ as depicted in \cref{fig:kc_epi}. While smaller eigenvalues are required for the episodic approach
compared to the asymptotic analysis in \cref{subsec:sim_data2e}, the maximum eigenvalue $\lambda_{\max}(\bm{A}_{\bm{\theta}})$ used in \cref{alg:itlearn} closely follow 
the expected $\mathcal{O}( \log(\nicefrac{1}{\sup_{t\geq0}\upsilon(t)}))$ behavior. Moreover, it can be directly seen that  \cref{alg:itlearn} offers a significant advantage over a direct reduction of the tracking error bound using the maximum eigenvalue $\lambda_{\max}(\bm{A}_{\bm{\theta}})$ without a compensation of the nonlinearity. Note that the sampling time $T_s$ necessary to achieve this behavior quickly decays as illustrated in \cref{fig:Ts_epi}. However, since it remains significantly larger than its theoretical bound $\underline{T}_s$, it remains 
in magnitudes which can be realized in practice. Therefore, \cref{alg:itlearn} provides an effective method for generating data, such that an arbitrary tracking error can be ensured when using a GP model for compensating unknown nonlinearities in systems of the form of \eqref{eq:sys}.\looseness=-1

\section{Conclusion}
\label{sec:conclusion}

This paper presents a novel, episodic approach for learning GP models in order
to ensure an arbitrarily high desired tracking accuracy using the GP to 
compensate unknown nonlinearities in linear systems. We first derive
a novel Bayesian prediction error bound for GP regression and demonstrate the 
straightforward computability of all required parameters. In order to establish
a straightforwardly interpretable connection between training data and prediction accuracy, 
we propose a kernel-dependent measure of data density and show that the prediction error
bound vanishes with increasing data density. We exploit the Bayesian error bounds to
derive a time-varying tracking error bound when using
the GP model to compensate unknown nonlinearities, and show that the tracking accuracy grows
with increasing data density. These theoretical results allow us to develop an 
episodic approach for learning a GP model, such that a desired tracking error bound
can be guaranteed. The effectiveness of our theoretical results is demonstrated in
several simulations.\looseness=-1

\bibliographystyle{IEEEtran}
\bibliography{example_paper}

\begin{thebibliography}{10}
\providecommand{\url}[1]{#1}
\csname url@samestyle\endcsname
\providecommand{\newblock}{\relax}
\providecommand{\bibinfo}[2]{#2}
\providecommand{\BIBentrySTDinterwordspacing}{\spaceskip=0pt\relax}
\providecommand{\BIBentryALTinterwordstretchfactor}{4}
\providecommand{\BIBentryALTinterwordspacing}{\spaceskip=\fontdimen2\font plus
\BIBentryALTinterwordstretchfactor\fontdimen3\font minus
  \fontdimen4\font\relax}
\providecommand{\BIBforeignlanguage}[2]{{%
\expandafter\ifx\csname l@#1\endcsname\relax
\typeout{** WARNING: IEEEtran.bst: No hyphenation pattern has been}%
\typeout{** loaded for the language `#1'. Using the pattern for}%
\typeout{** the default language instead.}%
\else
\language=\csname l@#1\endcsname
\fi
#2}}
\providecommand{\BIBdecl}{\relax}
\BIBdecl

\bibitem{Norgard2000}
P.~M. N{\o}rg{\aa}rd, O.~Ravn, N.~K. Poulsen, and L.~K. Hansen, \emph{{Neural
  Networks for Modelling and Control of Dynamical Systems - A Practicioner's
  Handbook}}.\hskip 1em plus 0.5em minus 0.4em\relax London: Springer, 2000.

\bibitem{Rasmussen2006}
C.~E. Rasmussen and C.~K.~I. Williams, \emph{{Gaussian Processes for Machine
  Learning}}.\hskip 1em plus 0.5em minus 0.4em\relax Cambridge, MA: The MIT
  Press, 2006.

\bibitem{Deisenroth2015}
M.~P. Deisenroth, D.~Fox, and C.~E. Rasmussen, ``{Gaussian Processes for
  Data-Effcient Learning in Robotics and Control},'' \emph{IEEE Transactions on
  Pattern Analysis and Machine Intelligence}, vol.~37, no.~2, pp. 408--423,
  2015.

\bibitem{Saveriano2017}
M.~Saveriano, Y.~Yin, P.~Falco, and D.~Lee, ``{Data-efficient control policy
  search using residual dynamics learning},'' in \emph{IEEE International
  Conference on Intelligent Robots and Systems}, 2017, pp. 4709--4715.

\bibitem{Nguyen-Tuong2009a}
D.~Nguyen-Tuong, M.~Seeger, and J.~Peters, ``{Local Gaussian Process Regression
  for Real Time Online Model Learning and Control},'' in \emph{Advances in
  neural information processing systems}, 2009, pp. 1193--1200.

\bibitem{Meier2016}
F.~Meier and S.~Schaal, ``{Drifting Gaussian processes with varying
  neighborhood sizes for online model learning},'' in \emph{Proceedings of the
  IEEE International Conference on Robotics and Automation}.\hskip 1em plus
  0.5em minus 0.4em\relax IEEE, 2016, pp. 264--269.

\bibitem{Lederer2021b}
A.~Lederer, A.~{Ordonez Conejo}, K.~Maier, W.~Xiao, J.~Umlauft, and S.~Hirche,
  ``{Gaussian Process-Based Real-Time Learning for Safety Critical
  Applications},'' in \emph{International Conference on Machine Learning},
  2021, pp. 6055--6064.

\bibitem{Yuan2020}
Z.~Yuan and M.~Zhu, ``{Communication-aware Distributed Gaussian Process
  Regression Algorithms for Real-time Machine Learning},'' in \emph{Proceedings
  of the American Control Conference}, 2020, pp. 2197--2202.

\bibitem{Lederer2022b}
A.~Lederer, Z.~Yang, J.~Jiao, and S.~Hirche, ``{Cooperative Control of
  Uncertain Multi-Agent Systems via Distributed Gaussian Processes},''
  \emph{IEEE Transactions on Automatic Control}, pp. 1--14, 2022.

\bibitem{Buisson-Fenet2019}
M.~Buisson-Fenet, F.~Solowjow, and S.~Trimpe, ``{Actively Learning Gaussian
  Process Dynamics},'' in \emph{Learning for Dynamics {\&} Control}, 2020, pp.
  1--11.

\bibitem{Capone2020c}
A.~Capone, G.~Noske, J.~Umlauft, T.~Beckers, A.~Lederer, and S.~Hirche,
  ``{Localized active learning of Gaussian process state space models},'' in
  \emph{Learning for Dynamics {\&} Control}, 2020, pp. 490--499.

\bibitem{Umlauft2018}
J.~Umlauft, L.~P{\"{o}}hler, and S.~Hirche, ``{An Uncertainty-Based Control
  Lyapunov Approach for Control-Affine Systems Modeled by Gaussian Process},''
  \emph{IEEE Control Systems Letters}, vol.~2, no.~3, pp. 483--488, 2018.

\bibitem{Hewing2020a}
L.~Hewing, J.~Kabzan, and M.~N. Zeilinger, ``{Cautious Model Predictive Control
  using Gaussian Process Regression},'' \emph{IEEE Transactions on Control
  Systems Technology}, vol.~28, no.~6, pp. 2736--2743, 2020.

\bibitem{Srinivas2012}
N.~Srinivas, A.~Krause, S.~M. Kakade, and M.~W. Seeger,
  ``{Information-Theoretic Regret Bounds for Gaussian Process Optimization in
  the Bandit Setting},'' \emph{IEEE Transactions on Information Theory},
  vol.~58, no.~5, pp. 3250--3265, 2012.

\bibitem{Kanagawa2018}
\BIBentryALTinterwordspacing
M.~Kanagawa, P.~Hennig, D.~Sejdinovic, and B.~K. Sriperumbudur, ``{Gaussian
  Processes and Kernel Methods: A Review on Connections and Equivalences},''
  pp. 1--64, 2018. [Online]. Available: \url{http://arxiv.org/abs/1807.02582}
\BIBentrySTDinterwordspacing

\bibitem{Chowdhury2017a}
S.~R. Chowdhury and A.~Gopalan, ``{On Kernelized Multi-armed Bandits},'' in
  \emph{Proceedings of the International Conference on Machine Learning}, 2017,
  pp. 844--853.

\bibitem{Scharnhorst2021}
\BIBentryALTinterwordspacing
P.~Scharnhorst, E.~T. Maddalena, Y.~Jiang, and C.~N. Jones, ``{Robust
  Uncertainty Bounds in Reproducing Kernel Hilbert Spaces: A Convex
  Optimization Approach},'' pp. 1--19, 2021. [Online]. Available:
  \url{http://arxiv.org/abs/2104.09582}
\BIBentrySTDinterwordspacing

\bibitem{Fiedler2021}
C.~Fiedler, C.~W. Scherer, and S.~Trimpe, ``{Practical and Rigorous Uncertainty
  Bounds for Gaussian Process Regression},'' in \emph{Proceedings of the AAAI
  Conference on Artificial Intelligence}, 2021.

\bibitem{Maddalena2020}
E.~T. Maddalena, P.~Scharnhorst, Y.~Jiang, and C.~N. Jones, ``{KPC:
  Learning-Based Model Predictive Control with Deterministic Guarantees},'' in
  \emph{Proceedings of the Conference on Learning for Dynamics {\&} Control},
  2020, pp. 1--12.

\bibitem{Umlauft2020}
J.~Umlauft and S.~Hirche, ``{Feedback Linearization Based on Gaussian Processes
  with Event-triggered Online Learning},'' \emph{IEEE Transactions on Automatic
  Control}, 2020.

\bibitem{Helwa2019}
M.~K. Helwa, A.~Heins, and A.~P. Schoellig, ``{Provably Robust Learning-Based
  Approach for High-Accuracy Tracking Control of Lagrangian Systems},''
  \emph{IEEE Robotics and Automation Letters}, vol.~4, no.~2, pp. 1587--1594,
  2019.

\bibitem{Lima2020}
G.~S. Lima, S.~Trimpe, and W.~M. Bessa, ``{Sliding Mode Control with Gaussian
  Process Regression for Underwater Robots},'' \emph{Journal of Intelligent and
  Robotic Systems: Theory and Applications}, vol.~99, no. 3-4, pp. 487--498,
  2020.

\bibitem{Maiworm2021}
M.~Maiworm, D.~Limon, and R.~Findeisen, ``{Online learning-based model
  predictive control with Gaussian process models and stability guarantees},''
  \emph{International Journal of Robust and Nonlinear Control}, pp. 1--28,
  2021.

\bibitem{Fiedler2021a}
C.~Fiedler, C.~W. Scherer, and S.~Trimpe, ``{Learning-enhanced robust
  controller synthesis with rigorous statistical and control-theoretic
  guarantees},'' in \emph{Proceedings of the IEEE International Conference on
  Decision and Control}, 2021.

\bibitem{Berkenkamp2016}
F.~Berkenkamp, A.~P. Schoellig, and A.~Krause, ``{Safe Controller Optimization
  for Quadrotors with Gaussian Processes},'' in \emph{Proceedings of the IEEE
  International Conference on Robotics and Automation}, 2016, pp. 491--496.

\bibitem{Marco2017}
A.~Marco, F.~Berkenkamp, P.~Hennig, A.~P. Schoellig, A.~Krause, S.~Schaal, and
  S.~Trimpe, ``{Virtual vs. Real: Trading off Simulations and Physical
  Experiments in Reinforcement Learning with Bayesian Optimization},'' in
  \emph{Proceedings of the IEEE International Conference on Robotics and
  Automation}, 2017, pp. 1557--1563.

\bibitem{Sui2015}
Y.~Sui, A.~Gotovos, J.~Burdick, and A.~Krause, ``{Safe Exploration for
  Optimization with Gaussian Processes},'' in \emph{Proceedings of The 32nd
  International Conference on Machine Learning}, 2015, pp. 997--1005.

\bibitem{Curi2020}
S.~Curi, F.~Berkenkamp, and A.~Krause, ``{Efficient Model-Based Reinforcement
  Learning through Optimistic Policy Search and Planning},'' in \emph{Advances
  in Neural Information Processing Systems}, 2020.

\bibitem{Curi2021}
S.~Curi, I.~Bogunovic, and A.~Krause, ``{Combining Pessimism with Optimism for
  Robust and Efficient Model-Based Deep Reinforcement Learning},'' in
  \emph{Proceedings of the International Conference on Machine Learning}, 2021,
  pp. 2254--2264.

\bibitem{Fossen2011}
T.~I. Fossen, \emph{{Handbook of Marine Craft Hydrodynamics and Motion
  Control}}.\hskip 1em plus 0.5em minus 0.4em\relax John Wiley {\&} Sons, 2011.

\bibitem{Yu2015}
H.~Yu, S.~Huang, G.~Chen, Y.~Pan, and Z.~Guo, ``{Human-Robot Interaction
  Control of Rehabilitation Robots with Series Elastic Actuators},'' \emph{IEEE
  Transactions on Robotics}, vol.~31, no.~5, pp. 1089--1100, 2015.

\bibitem{Lederer2019}
A.~Lederer, J.~Umlauft, and S.~Hirche, ``{Uniform Error Bounds for Gaussian
  Process Regression with Application to Safe Control},'' in \emph{Advances in
  Neural Information Processing Systems}, 2019.

\bibitem{Capone2019}
A.~Capone and S.~Hirche, ``{Backstepping for Partially Unknown Nonlinear
  Systems Using Gaussian Processes},'' \emph{IEEE Control Systems Letters},
  vol.~3, no.~2, pp. 416--421, 2019.

\bibitem{Skogestad2005}
S.~Skogestad and I.~Postlethwaite, \emph{{Multivariable Feedback Control:
  Analysis and Design}}, 2nd~ed.\hskip 1em plus 0.5em minus 0.4em\relax New
  York, NY: John Wiley {\&} Sons, 2005.

\bibitem{Perko2006}
L.~Perko, \emph{{Differential Equations and Dynamical Systems}}, 3rd~ed.\hskip
  1em plus 0.5em minus 0.4em\relax Springer, 2006.

\bibitem{Dhiman2021}
V.~Dhiman, M.~J. Khojasteh, M.~Franceschetti, and N.~Atanasov, ``{Control
  Barriers in Bayesian Learning of System Dynamics},'' \emph{IEEE Transactions
  on Automatic Control}, pp. 1--16, 2021.

\bibitem{VanderVaart2011}
A.~van~der Vaart and H.~van Zanten, ``{Information Rates of Nonparametric
  Gaussian Process Methods},'' \emph{Journal of Machine Learning Research},
  vol.~12, pp. 2095--2119, 2011.

\bibitem{Capone2021b}
A.~Capone, A.~Lederer, and S.~Hirche, ``{Gaussian Process Uniform Error Bounds
  with Unknown Hyperparameters for Safety-Critical Applications},'' in
  \emph{Proceedings of the International Conference on Machine Learning}, 2022,
  pp. 2609--2624.

\bibitem{Kuleshov2018}
V.~Kuleshov, N.~Fenner, and S.~Ermon, ``{Accurate uncertainties for deep
  learning using calibrated regression},'' in \emph{Proceedings of the
  International Conference on Machine Learning}, 2018, pp. 4369--4377.

\bibitem{Shalev-Shwartz2013}
S.~Shalev-Shwartz and S.~Ben-David, \emph{{Understanding Machine Learning: From
  Theory to Algorithms}}.\hskip 1em plus 0.5em minus 0.4em\relax New York, NY:
  Cambridge University Press, 2013.

\bibitem{Dudley1967}
R.~M. Dudley, ``{The Sizes of Compact Subsets of Hilbert Space and Continuity
  of Gaussian Processes},'' \emph{Journal of Functional Analysis}, vol.~1,
  no.~3, pp. 290--330, 1967.

\bibitem{Grunewalder2010}
S.~Gr{\"{u}}new{\"{a}}lder, J.-Y. Audibert, M.~Opper, and J.~Shawe-Taylor,
  ``{Regret Bounds for Gaussian Process Bandit Problems},'' \emph{Journal of
  Machine Learning Research}, vol.~9, pp. 273--280, 2010.

\bibitem{Talagrand1994}
M.~Talagrand, ``{Sharper Bounds for Gaussian and Empirical Processes},''
  \emph{The Annals of Probability}, vol.~22, no.~1, pp. 28--76, 1994.

\bibitem{Ghosal2006}
S.~Ghosal and A.~Roy, ``{Posterior Consistency of Gaussian Process Prior for
  Nonparametric Binary Regression},'' \emph{The Annals of Statistics}, vol.~34,
  no.~5, pp. 2413--2429, 2006.

\bibitem{Laurent2000}
B.~Laurent and P.~Massart, ``{Adaptive Estimation of a Quadratic Functional by
  Model Selection},'' \emph{The Annals of Statistics}, vol.~28, no.~5, pp.
  1302--1338, 2000.

\bibitem{Gershgorin1931}
S.~A. Gershgorin, ``{Ueber die Abgrenzung der Eigenwerte einer Matrix},''
  \emph{Bulletin de l'Academie des Sciences de l'URSS. Classe des sciences
  mathematiques et na}, no.~6, pp. 749--754, 1931.

\bibitem{Shekhar2018}
S.~Shekhar and T.~Javidi, ``{Gaussian Process Bandits with Adaptive
  Discretization},'' \emph{Electronic Journal of Statistics}, vol.~12, pp.
  3829--3874, 2018.

\bibitem{Vivarelli1998}
F.~Vivarelli, ``{Studies on the Generalisation of Gaussian Processes and
  Bayesian Neural Networks},'' Ph.D. dissertation, Aston University, 1998.

\bibitem{Khalil2002}
H.~K. Khalil, \emph{{Nonlinear Systems}}, 3rd~ed.\hskip 1em plus 0.5em minus
  0.4em\relax Upper Saddle River, NJ: Prentice-Hall, 2002.

\end{thebibliography}

\vspace{-1.0cm}
\begin{IEEEbiography}[{\includegraphics[width=1in,height=1.25in,clip,keepaspectratio]{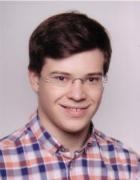}}]{Armin Lederer}
	(S'20) 
	received the B.Sc. and M.Sc. degree in electrical engineering and information technology from the Technical University of Munich, Germany, in 2015 and 2018, respectively.
	Since June 2018, he has been a PhD student at the Chair of Information-oriented Control, Department of Electrical and Computer Engineering at the Technical University of
	Munich, Germany. His current research interests
	include the stability of data-driven control systems and machine learning in closed-loop systems.
\end{IEEEbiography}
\vspace{-1.0cm}
\begin{IEEEbiography}[{\includegraphics[width=1in,height=1.25in,clip,keepaspectratio]{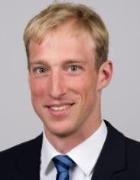}}]{Jonas Umlauft}
	(S’14) received the B.Sc. and M.Sc. degree in electrical engineering and information technology from the Technical University of Mu- nich, Germany, in 2013 and 2015, respectively. His Master’s thesis was completed at the Computational and Biological Learning Group at the University of Cambridge, UK. Since May 2015, he has been a PhD student at the Chair of Information-oriented Control, Department of Electrical and Computer Engineering at the Technical University of Munich, Germany. His current research interests
	include the stability of data-driven control systems and system identification based on Gaussian processes.
\end{IEEEbiography}
\vspace{-1.0cm}
\begin{IEEEbiography}[{\includegraphics[width=1in,height=1.25in,clip,keepaspectratio]{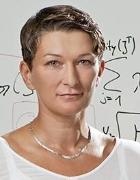}}]{Sandra Hirche}
	(M'03--SM'11--F'20) 
	received the Dipl.-Ing degree in aeronautical engineering from the Technical University of Berlin, Berlin, Germany, in 2002, and the Dr. Ing. degree in electrical engineering from the Technical University of Munich, Munich, Germany, in 2005. From 2005 to 2007, she was awarded a Post-doctoral scholarship from the Japanese Society for the Promotion of Science at the Fujita Laboratory, Tokyo Institute of Technology, Tokyo, Japan. From 2008 to 2012, she was an Associate Professor with the Technical University of Munich. Since 2013, she has served as Technical University of Munich Liesel Beckmann Distinguished Professor and has been with the Chair of Information-Oriented Control, Department of Electrical and Computer Engineering, Technical University of Munich. She has authored or coauthored more than 150 papers in international journals, books, and refereed conferences. Her main research interests include cooperative, distributed, and networked control with applications in human--machine interaction, multirobot systems, and general robotics. 
	
	Dr. Hirche has served on the editorial boards of the IEEE Transactions on Control of Network Systems, the IEEE Transactions on Control Systems Technology, and the IEEE Transactions on Haptics. She has received multiple awards such as the Rohde \& Schwarz Award for her Ph.D. thesis, the IFAC World Congress Best Poster Award in 2005, and -- together with students -- the 2018 Outstanding Student Paper Award of the IEEE Conference on Decision and Control as well as Best Paper Awards from IEEE Worldhaptics and the IFAC Conference of Manoeuvring and Control of Marine Craft in 2009.
\end{IEEEbiography}



\vfill


%
%
%
%


\end{document}